\documentclass[prd,preprintnumbers,twocolumn,notitlepage]{revtex4-2}

\usepackage{multirow}
\usepackage{siunitx}

\usepackage{tikz}
\usetikzlibrary{decorations.markings, arrows}

\usepackage{graphicx,times}
\usepackage{latexsym}
\usepackage{mathtools}
\usepackage{amsmath,amssymb,amsbsy,amsfonts}
\usepackage{array}
\usepackage{bm}
\usepackage{graphics}
\usepackage{mathrsfs}
\usepackage{xcolor}
\usepackage{cancel}
\usepackage[normalem]{ulem}

\usepackage[unicode, pdfprintscaling=None]{hyperref}
\hypersetup{
    colorlinks,
    linkcolor={red!50!black},
    citecolor={blue!50!black},
    urlcolor={blue!80!black}
}

\usepackage{microtype}

\usepackage[capitalise]{cleveref}

\usepackage{booktabs}
\usepackage{ctable}
\usepackage{xcolor}
\usepackage{makecell}

\newcommand{\RR}{{\mathbb{R}}}

\renewcommand\>{\rangle}
\newcommand\<{\langle}

\DeclareMathOperator\Tr{Tr}
\newcommand\Cfg{\mathcal{C}}

\usepackage{relsize}

\usepackage[acronyms,smallcaps,nohypertypes={acronym},nopostdot,style=super,nonumberlist, toc]{glossaries}
\glsenableentrycount
\makeglossaries  
\setacronymstyle{long-sc-short}

\newcommand\ac[1]{\gls{#1}}
\newcommand\acp[1]{\glspl{#1}}

\newcommand\Lmax{L_{\text{max}}}
\newcommand\Lmin{L_{\text{min}}}
\newcommand\xdagger{\vphantom{\dagger}}

\renewcommand\bf\mathbf

\newcommand\Id{\mathbb{I}}

\newcommand\HbSp{\mathcal{H}}
\newcommand\vac{\varnothing}

\newcommand\Ztotal{Z'}
\newcommand\Zworm{Z_{w}}

\newcommand\CartanH{h}
\newcommand\laO{\mathfrak{o}}

\newcommand\Jpp{J_2}
\newcommand\Jt{J_1}
\newcommand\Jhop{J_2'}
\newcommand\nMax{n_{\text{max}}}

\newcommand\HSSing{\HbSp_{x}^{(0)}}
\newcommand\HSFund{\HbSp_{x}^{(1)}}

\graphicspath{{./plots/}{./fig/}}

\newacronym{WF}{wf}{Wilson-Fisher}
\newacronym{AF}{af}{asymptotically free}

\newacronym{RG}{rg}{renormalization group}

\newacronym[longplural={conformal field theories}]{CFT}{cft}{conformal field theory}
\newacronym[longplural={lattice field theories}]{LFT}{lft}{lattice field theory}
\newacronym[longplural={effective field theories}]{EFT}{eft}{effective field theory}
\newacronym[longplural={quantum field theories}]{QFT}{qft}{quantum field theory}

\newacronym{JLP}{jlp}{Jordan-Lee-Preskill}

\newacronym{BBN}{bbn}{big bang nucleosynthesis}

\newacronym{LEC}{lec}{low-energy constant}

\newacronym{QCD}{qcd}{quantum chromodynamics}
\newacronym{MC}{mc}{Monte Carlo}

\newacronym{IR}{ir}{infrared}
\newacronym{UV}{uv}{ultraviolet}

\newacronym{QED}{qed}{quantum electrodynamics}

\newacronym{SNR}{snr}{signal-to-noise ratio}

\newacronym{NLSM}{nlsm}{nonlinear sigma model}

\newacronym{CL}{cl}{Complex Langevin}

\newacronym{CSA}{csa}{Cartan subalgebra}

\newacronym{SSB}{ssb}{spontaneous symmetry breaking}

\newacronym{AFQMC}{afqmc}{auxiliary field quantum Monte Carlo}
\newacronym{iHMC}{ihmc}{imaginary-mass Hybrid Monte Carlo}

\newacronym{MCMC}{mcmc}{Markov Chain Monte Carlo}

\begin{document}

\title{\texorpdfstring{Qubit regularized $O(N)$ nonlinear sigma models}{Qubit regularized $O(N)$ nonlinear sigma models}}
\author{Hersh Singh}
\preprint{IQuS@UW-21-021,INT-PUB-22-007}
\email{hershsg@uw.edu}
\affiliation{InQubator for Quantum Simulation (IQuS), Department of Physics, University of Washington, Seattle, Washington 98195-1550, USA}

\begin{abstract}
Motivated by the prospect of quantum simulation of quantum field theories, we formulate the $O(N)$ nonlinear sigma model as a ``qubit'' model with an $(N+1)$-dimensional local Hilbert space at each lattice site. Using an efficient worm algorithm in the worldline formulation, we demonstrate that the model has a second-order critical point in $(2+1)$ dimensions, where the continuum physics of the nontrivial $O(N)$ Wilson-Fisher fixed point is reproduced. We compute the critical exponents $\nu$ and $\eta$ for the $O(N)$ qubit models up to $N=8$, and find excellent agreement with known results in literature from various analytic and numerical techniques for the $O(N)$ Wilson-Fisher universality class. Our models are suited for studying $O(N)$ nonlinear sigma models on quantum computers up to $N=8$ in $d=2,3$ spatial dimensions.
\end{abstract}

\maketitle

\section{Introduction}
\label{sec:intro}

Quantum field theory is the best known framework for a fundamental description of nature, as evidenced by the success of the standard model of particle physics, and describes long-distance features of condensed matter systems close to criticality.
While lattice Monte Carlo computations can be used for static properties in cases where the sign problem can be controlled, studying non-perturbative and non-equilibrium properties of generic \acp{QFT} remains an important yet daunting task.  

With advancements in quantum technology, there have been significant advances in making digital and analog quantum platforms available for widespread use in scientific applications. 
The hope that quantum computers might solve the issues which plague classical lattice Monte Carlo computations of \acp{QFT}
\cite{feynman_simulating_1982, benioff_computer_1980}
has motivated a surge of activity with the goal of quantum simulation of \acp{QFT}, with
\cite{preskill_simulating_2018, preskill_quantum_2018, dalmonte_lattice_2016, banuls_simulating_2019, banuls_review_2019, alexeev_quantum_2020}.

Among the various quantum systems which one might like to study, \acp{QFT} are unique in that many, drastically different, microscopic descriptions can yield the same universal physics as one takes the continuum limit \cite{kogut_introduction_1979, wilson_renormalization_1983, wilson_renormalization_1974}.
This freedom in choosing a microscopic description has always been used in constructing better actions for classical lattice field theory computations \cite{symanzik_continuum_1983b, symanzik_continuum_1983a}.
However, it becomes even more relevant when one starts consider the wide variety of quantum hardware. The best regularization for a given \ac{QFT} will certainly depend on the type of quantum hardware at hand.

A reason to suspect that traditional lattice regularizations of bosonic field theories might be unsuitable for quantum computers is the issue of local Hilbert spaces. Traditional lattice regularizations of bosonic theories, including gauge theories such as \ac{QCD}, have infinite-dimensional local Hilbert spaces
\cite{kogut_hamiltonian_1975}, where each site has an infinite-dimensional harmonic-oscillator-like degree of freedom.
However, most quantum technologies being pursued today realize finite-dimensional qubit (or qudit) systems.
Using a traditional lattice regularization, we are forced to truncate the local field values so that the local infinite-dimensional Hilbert space maps to, say, a $n$-dimensional Hilbert space of the qubits. Even though, in principle, we would recover the traditional model as $n \to \infty$, this introduces an additional systematic which we need to be careful about. 
Much of the recent effort in the community has been dedicated to finding sensible truncations of the local Hilbert space \cite{zohar_formulation_2015a, zohar_digital_2017, nuqscollaboration_ensuremath_2019, martinez_realtime_2016a, klco_digitization_2019, ji_gluon_2020, haase_resource_2021, bruckmann_nonlinear_2019, bronzan_explicit_1985, bender_gauge_2020, alexandru_spectrum_2021}.

On the other hand, it might be unnecessary to introduce this systematic. Indeed, it is well-known, since Wilson's \ac{RG} \cite{wilson_renormalization_1983, wilson_renormalization_1974, kogut_introduction_1979}, that continuum \acp{QFT} with infinite-dimensional local Hilbert spaces can arise close to critical points of quantum many-body systems with finite-dimensional local Hilbert spaces.
For example, the quantum transverse field Ising model, with just spin-$1/2$ local degrees of freedom, famously reproduces the continuum $\phi^4$ theory near a critical point in $d \geq 2$ spatial dimensions.
Many other \acp{QFT} are also known to arise from continuum limits of such spin models near critical points.
From the point of view of quantum simulation of \acp{QFT}, it has become an especially relevant and interesting question to find nontraditional ways of obtaining continuum \acp{QFT}, which use the available qubits efficiently.
To emphasize this, in Ref.~\cite{singh_qubit_2019}, we used the term \emph{qubit regularization} of a \ac{QFT} to denote such a regularization with a finite-dimensional local Hilbert space, which reproduces a given \ac{QFT} close to a quantum critical point.
This approach to finding such unconventional regularizations for \acp{QFT} is being pursued actively
\cite{zhou_spacetime_2021, bhattacharya_qubit_2021, liu_qubit_2022, singh_qubit_2019,roy_quantum_2021,roy_quantum_2019}.

The goal of this work is to apply these ideas to the $O(N)$ nonlinear sigma models in $D\geq3$ spacetime dimensions and construct a lattice regularization in the Hamiltonian framework with a finite-dimensional local Hilbert space.
The $O(N)$ \ac{NLSM} is a \ac{QFT}, perturbatively defined by the continuum Lagrangian,
\begin{align}
  \mathcal S[\vec \phi] = -\frac1{g^2} \int d^{D}x\ \partial_\mu \vec \phi \cdot \partial^\mu \vec \phi,
\end{align}
where $\vec \phi(x) \in \RR^N$ is an  $N$-component bosonic scalar field with the constraint $|\vec \phi(x) | =1$.
The $O(N)$ \acp{NLSM}, for various $N$ and dimensions, have a rich phenomenology  and many physical applications. In spacetime dimensions $D=3$, they find applications in describing magnetism \cite{zinn-justin_quantum_2021}, the superfluid transition in Helium-4 \cite{campostrini_critical_2006} and the spontaneous breakdown of chiral symmetry in two-flavor \ac{QCD} with light quarks \cite{pisarski_remarks_1984}.
In $D=2$ spacetime dimensions, the $O(N)$ \ac{NLSM} are very special owing to their integrability \cite{zamolodchikov_factorized_1979}, offer an excellent toy model for \ac{QCD}, displaying features such as asymptotic freedom, dynamical mass generation and dimensional transmutation, and often arise as low-energy effective theories of spin chains and ladders \cite{affleck_critical_1987a, haldane_continuum_1983, haldane_nonlinear_1983}. 

On the lattice, the continuum $O(N)$ \ac{NLSM} has been extensively studied using the lattice regulated action on a $D$-dimensional Euclidean lattice,
\begin{align}
    \mathcal S[\vec \phi] = -\beta \sum_{\< xy \>} \vec \phi_{x} \cdot \vec \phi_{y}
  \label{eq:ON-trad}
\end{align}
where $\beta$ is a coupling,
$\vec \phi_{x} \in \RR^{N}$ is an $N$-component real-valued field with the constraint $|\vec \phi_x|^2 = 1$ for all sites $x$, and 
the sum runs over all nearest-neighbor lattice sites $x, y$.
At a finite critical value of the coupling $\beta = \beta_c$, this model is known to have a second-order phase transition in $D=3,4$ dimensions.
For $D=4$ the critical point is described by the trivial Gaussian fixed point. More interestingly, for $D=3$, the critical point is described by the nontrivial \ac{WF} fixed point \cite{wilson_critical_1972}.
In $D=2$, this model is asymptotically free with a critical point at $\beta_c \to \infty$.

In Ref.~\cite{singh_qubit_2019}, the authors showed that a simple qubit regularization for the $O(3)$ \ac{NLSM} can be constructed using just two qubits per lattice site, which reproduces the continuum physics for $D=3,4$ spacetime dimensions.
Similar models were also used earlier in Ref.~\cite{cecile_modeling_2008} to model pion physics, and in Refs.~\cite{banerjee_conformal_2018, banerjee_conformal_2019} for the $O(2)$ and $O(4)$ models to study sectors of large charges, which have a signal-to-noise ratio problem in the conventional formulation of \cref{eq:ON-trad}.

In this work, we extend these earlier results to show that the construction can generalized to the $O(N)$ case for arbitrary $N$, with an $(N+1)$-dimensional local Hilbert space. From a Hilbert space perspective, these qubit models are the simplest and most economical, since they only use the smallest two representations of $O(N)$. We emphasize that while it is easy to write down many Hamiltonian models with an $O(N)$ symmetry, it is also important to show that there is a quantum critical point in the right universality class for the qubit model to work as a regularization of a \ac{QFT}, which in general can be a hard problem. One attractive feature of our proposed model is that there is a nice worldline representation which is amenable to an efficient worm algorithm, which allows us to use classical lattice Monte Carlo computations to show that the model indeed has a critical point in the $O(N)$ \ac{WF} universality class.

Here, we confine ourselves to $D\geq 3$ spacetime dimensions, and in particular perform numerical computations for $D=3$, where the $O(N)$ \ac{NLSM} are controlled by the nontrivial \ac{WF} fixed point. However, the $D=2$ case is especially interesting as well, due to asymptotic freedom for $N\geq 3$.
Even though it is, in principle, possible to obtain asymptotic freedom from qubit models, it seems to be nontrivial to find the right critical point.
Recently, in Ref.~\cite{bhattacharya_qubit_2021}, the authors provided strong numerical evidence to support that in fact this can be done using just two qubits per lattice site. The continuum limit $O(3)$ \ac{NLSM} in $1+1$ dimensions is achieved with a fixed two-qubit local Hilbert space.
In our preliminary investigations, the $O(N)$ qubit models proposed in this work do not seem to have the right continuum limit in $1+1$ dimensions \cite{zhou_spacetime_2021} with a finite-dimensional local Hilbert space. It remains an interesting open question whether there are $O(N)$ qubit models in $1+1$ dimensions which show asymptotic freedom.

This paper is organized as follows.
In \cref{sec:qubit-model}, we describe the construction of a Hamiltonian model with $O(N)$ symmetry. In \cref{sec:algorithm}, we show how to construct a worldline representation and develop an efficient worm-algorithm for it.
In \cref{sec:criticality}, we perform computations for the $O(N)$ qubit models for various $N$ and compare our results for the critical point with known results in literature. Finally, in \cref{sec:conclusions}, we present our conclusions and comment on related future work.

\section{The qubit model}
\label{sec:qubit-model}

In this section, we would like to write down a simple $O(N)$-invariant Hamiltonian, which we will show to have a critical point in the $O(N)$ \ac{WF} universality class. The smallest irreps (irreducible representations) of $O(N)$ are the trivial singlet (1-dimensional) and the vector ($N$-dimensional) representations. We will construct a Hamiltonian model using only these irreps for the local Hilbert space.

Let $|i\>$ (where $i=1,\dotsc,N$ for $N$ even, and $i=0, 1, \dotsc, N-1$ for $N$ odd) be a basis for the fundamental representation of the $O(N)$ group.
We take this to be the ``Cartesian basis,'' such that rotations acts on the basis vector $|i\>$ as
\begin{align}
  |i\> \to R_{ij}|j\>
\end{align}
where $R \in O(N)$ is an $N \times N$ matrix satisfying $R^T R = \Id$.

It will be more convenient for us to work in a basis where the states diagonalize a \ac{CSA} of the $\laO(N)$ Lie algebra.
Physically, such a basis consists of states with well-defined $O(N)$ charges.
To make it easier to discuss both even and odd $N$ at once, we shall define an integer $n$ such that $N \equiv 2n$ for $N$ even, and $N \equiv 2n+1$ for $N$ odd.
Let us also define a set of integers $K_N$ such that $K_N = \{0, \pm1, \dotsc, \pm n\}$ for $N$ odd, and $K_N = \{\pm1, \dotsc, \pm n\}$ for $N$ even.
We define $\CartanH_i$ ($i=1, \dotsc, n$) as the generator of rotations in the $(2i-1,2i)$ plane,
\begin{align}
  \CartanH_i = i(E_{2i,2i-1} - E_{2i-1,2i}) \quad\text{for}\quad i=1,\dotsc,n,
  \label{eq:cartan-matrix-elements}
\end{align}
where $E_{i,j}$ is an $N \times N$ matrix with the $(i,j)$th element as one, and all other matrix elements zero. 
Clearly, the $\CartanH_i$ commute with each other,
\begin{align}
  [\CartanH_i, \CartanH_j ] = 0 \quad \text{for all}\quad i,j = 1,\dotsc, n.
\end{align}
and form a basis for the \ac{CSA} of the $\laO(N)$  lie algebra. 

Since all the Cartan generators commute with each other, they can be simultaneously diagonalized.
Writing the Cartan generators as a vector $\vec h \equiv (h_1, \dotsc, h_n)$, 
we let the simultaneous eigenvectors of the \ac{CSA} be labeled by an $n$-dimensional vector $\vec{q}^{\,k} \equiv (q^{k}_{1}, \dotsc, q^{k}_{n})$ of eigenvalues, such that
\begin{align}
  h_j | \vec q^{\,k} \> = q^k_{j}|\vec q^{\,k} \>, \quad k \in K_N.
\end{align}
Explicitly, in terms of the Cartesian basis vectors $|i\>$, these eigenvectors are given by
\begin{align}
  | \vec q^{\,\pm k} \> =
  \begin{cases}
    \frac{1}{\sqrt{2}}\left(| 2k-1 \> \pm i | 2k \>\right)\quad &\text{for } k = 1, \dotsc, n,\\
    |0\> &\text{for $k=0$ (if $N$ is odd)}
  \end{cases}
          \label{eq:spherical-basis}
\end{align}
with the eigenvalues
\begin{align}
  (q^{\,\pm k})_{j} = \begin{cases}
                                                               \pm \delta_{kj} \quad &\text{for } k=1,\dotsc, n,  \\
                                                               0 &\text{for } k=0.
                                                               \end{cases}
\end{align}

As mentioned earlier, this basis is convenient for us since each of the states $|\vec q^{\,k}\>$ have well-defined $O(N)$ charges. That is,
under an $O(N)$ rotation generated by the Cartan generators, parameterized by $\vec \theta =(\theta_1, \dotsc, \theta_n)$, 
the state $| \vec q^{\,k} \>$ transforms with just an overall phase,
\begin{align}
  | \vec q^{\,k } \> \to e^{i \vec \theta\cdot \vec \CartanH} | \vec q^{\, k} \> = e^{i \vec \theta \cdot \vec q^{\,k}} | \vec q^{\, k} \>.
\end{align}
In this sense, we say that the state $| \vec q^{\, k} \>$ has the $O(N)$ charge $\vec q^{\,k}$.
We note that, if $N$ is odd, then we also have a state $|\vec q_0\>$ which has zero charge,
\begin{align}
  \vec q_0 = (0,\dotsc,0).
\end{align}
We may call the basis given by \cref{eq:spherical-basis} as the ``spherical basis,'' in analogy with rotations in three spatial dimensions. 

We can now construct a qubit model for the $O(N)$ \ac{NLSM}, 
The full Hilbert space $\HbSp$ is the tensor product of local Hilbert spaces $\HbSp_x$,
\begin{align}
  \HbSp = \otimes_{x} \HbSp_x,
\end{align}
where $x$ varies over all lattice sites.
We define the local Hilbert space $\HbSp_x$ to be an $(N+1)$-dimensional vector space, written as a direct sum
\begin{align}
  \HbSp_x = \HbSp_x^{(0)} \oplus \HbSp_x^{(1)}.
\end{align}
where $\HbSp_x^{(0)}$ is the one-dimensional singlet representation, and $\HbSp_x^{(1)}$ is the $N$-dimensional fundamental representation.

Let the singlet Hilbert space $\HSSing$ be spanned by a normalized vector $|\vac \>$.
We will think of this state as the Fock vacuum.
For the fundamental representation $\HSFund$, we use the spherical basis $|\vec q^{\, k}\> $ with  $k\in K_N$.
We interpret the state $|\vec q^{\,k}\> \in \HSFund$ as a ``single-particle'' state with charge $\vec q^{\,k}$.
We can take this interpretation further and define creation and annihilation operators at each lattice site $x$, which can respectively be written as
\begin{align}
  c_{x, k }^\dagger &= |\vec q^{\,k}\>_x \< \vac |_x, \quad\text{and}  \\
  c_{x, k } &= |\vac\>_x\<\vec q^{\,k}|_x,
\end{align}
where $k \in K_N$.
We think of the operators $c^{\dagger}_{x, k }$ and $c_{x, k}$ as creating and annihilating particles of charge $\vec q^{\,k}$ at the site $x$, respectively.

Note that applying a creation operation twice on a site will give zero.
In other words, we do not allow more than one particle at a site.
So, in a sense, these particles obey a kind of Pauli exclusion principle.
However, these operators are not truly fermionic since they commute at different lattice sites.
Such particles are often called hard-core bosons in condensed matter literature.

Finally, we can now write down a simple $O(N)$ invariant Hamiltonian with nearest-neighbor terms,
\begin{align}
  H &= \Jt \sum_{\substack{x\\k\in K_N}} c^{\dagger}_{x, k} c_{x, k}^{\xdagger}
      - \Jpp \sum_{\substack{\< xy \>\\k\in K_N} } \left( c^{\dagger}_{x, k} c_{y, k}^{\xdagger}  + c_{y, k}^\dagger c_{x, k}^{\xdagger} \right)  \nonumber\\
      &\quad - \Jhop \sum_{\substack{\< xy \>\\k\in K_N} } \left( c^{\dagger}_{x, k} c_{y, -k}^\dagger + c_{x,k} c_{y,-k} \right)
      \label{eq:ON-qubit-ham}
\end{align}
where $x$ runs over all lattice sites on a $d$-dimensional spatial lattice, $\< x y\>$ runs over all nearest neighbor sites, and $J_1,J_2, J_2'$ are couplings.
Setting $N = 3$, we can see that this is exactly the $O(3)$ qubit model that was constructed in Ref.~\cite{singh_qubit_2019}.

To provide some intuition for the Hamiltonian in \cref{eq:ON-qubit-ham}, we note that there are three terms. The first one is simply an on-site term which makes the single-particle states costly for $J_1 > 0$.
The second term is a hopping term for particles of a given charge. The third term allows for a particle-antiparticle pair (of charges $\vec q_{k}$ and $-\vec q_{k}$) to be created out of the Fock vacuum, or for such a particle-antiparticle pair to be annihilated into the vacuum.

\begin{figure}[t]
    \centering
    \includegraphics[]{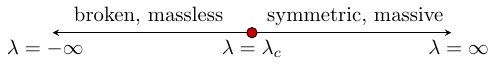}
    \caption{The zero temperature phase diagram of our qubit Hamiltonian of \cref{eq:ON-qubit-ham} in $d=2,3$ spatial dimensions, with $J_2 = J_2' > 0 $ and $\lambda = J_1/J_2$. As $\lambda \to \infty$, singlets dominate and the system is in a massive phase. However, as $\lambda \to -\infty$, the ground state is dominated by the $O(N)$ particles and $O(N)$ symmetry must be spontaneously broken, giving rise to massless goldstone modes. We expect there to be a second-order phase transition at some intermediate $\lambda = \lambda_c$. }
    \label{fig:phasediag}
\end{figure}

Of course, this is just one of the infinitely-many $O(N)$-invariant Hamiltonians we may write down. Here, the idea is to write down the simplest Hamiltonian  with the right symmetries which can be shown to have the right continuum limit.
Indeed, we can heuristically argue that the model given in \eqref{eq:ON-qubit-ham} should have a phase transition in $d \geq 2$ spatial dimensions.
For simplicity, we take $\Jpp = \Jhop > 0$,
and let $\lambda \equiv \Jt/\Jpp$. The expected phase diagram for this model is shown in \cref{fig:phasediag}.
First we consider the limit in which $\lambda$ is large and positive.
In this limit, the fundamental states are energetically suppressed due to the $\Jt$ term in the Hamiltonian.
The ground state of the theory is therefore dominated by the singlets.
In the $\lambda \to +\infty $ limit, the ground state is just the $O(N)$ symmetric Fock vacuum $| \Omega \> = | \vac \vac \cdots \vac \>$.
Making the $\lambda$ finite (but still large and positive) introduces fluctuations in the system in the form of single-particle states.
Since the lowest excited states consist of an energetically costly $N$-tuplet of the $O(N)$ particles, the system should be gapped in this regime.

On the other hand, if we take $\lambda$ large and \emph{negative}, the fundamental states $|\vec q^{\, k}\>$ start to dominate the ground state.
In the $\lambda \to -\infty$ limit, the $O(N)$ symmetry of the vector particles suggests that ground state is $N$-fold degenerate.
However, for $d\geq2$ in the thermodynamic limit, we expect this $O(N)$ symmetry to be spontaneously broken, giving rise to Goldstone modes. The system must thus be gapless in this regime.

Therefore, as we tune $\lambda$ from large and positive to large and negative, at some critical $\lambda = \lambda_c$, the system must undergo a phase transition associated with the spontaneous breaking of the $O(N)$ symmetry.
If the phase transition is second order, then a continuum \ac{QFT} should emerge at the critical point, according to the ideas of Wilson's \ac{RG}.
In $d \geq 2$, we expect the emergent \ac{QFT} to be precisely the $O(N)$ \ac{NLSM}, which is controlled by the \ac{WF} fixed point in $d=2$ and the Gaussian (free) fixed point in $d=3$ (with a marginally irrelevant parameter).

Our aim in this work to is to develop the worldline formulation and worm algorithm to numerically verify this heuristic argument.
We do this in the next section.
If the continuum \ac{QFT} is indeed the $O(N)$ \ac{NLSM}, then this opens up the possibility of studying the physics of the continuum $O(N)$ \ac{NLSM} using such qubit models, both on classical and quantum computers.

\section{Worldline formulation and the worm algorithm}
\label{sec:algorithm}

\subsection{Worldline formulation}

Starting from the Hamiltonian in \cref{eq:ON-qubit-ham}, we can construct a worldline formulation which would be amenable to Monte Carlo computations using a worm algorithm.  
To summarize the result of this section, 
the worldline formulation for the $O(2n)$ model gives us a model of non-intersecting oriented worldlines (in spacetime) with $n$ colors (corresponding to the various charges), with the loop weights symmetric across the $n$ colors.  For the $O(2n+1)$ model, we get a model of $n$ oriented  worldlines and an additional unoriented worldline, corresponding to the state of zero charge.  

Such a construction is well-known and we merely sketch the procedure (see, for example, Refs.~\cite{singh_qubit_2019, syljuasen_quantum_2002a, singh_fewbody_2019}).
We begin with the partition function
\begin{align}
  Z = e^{-\beta H}.
\end{align}
To get the worldline representation, it is convenient to introduce a spacetime lattice by treating $\beta$ as imaginary time and splitting $\beta$ into $L_T$ pieces with $\beta = \varepsilon L_T$,
\begin{align}
    Z &=  \Tr \left[ e^{-\varepsilon H} \cdots e^{-\varepsilon H} \right] .
\end{align}
We write the Hamiltonian as $H = H_1 + H_2$, where $H_1$ is a sum over single-site terms, and $H_2$ is a sum over nearest-neighbor terms.  In the local $|\vac\>, |\vec q^{\, k}\>$ basis, $H_1$ is diagonal, while $H_2$ has off-diagonal terms.   
To exploit this fact, we approximate the Hamiltonian as
\begin{align}
    e^{-\varepsilon (H_1 + H_2)} = e^{-\varepsilon H_1} \left( 1 - \varepsilon H_2  + O(\varepsilon^2) \right).
\end{align}
which is valid for small enough $\varepsilon$.
Now, we can evaluate the trace in our local basis and insert a complete set of states after each time step, which results in a sum over worldline configurations,
\begin{align}
    Z &=  \sum_{\Cfg} W[\mathcal{C}] \label{eq:Z}
\end{align}
where $\Cfg$ is a worldline configuration on a space-time lattice (periodic in the time direction), and $W[\Cfg]$ is the weight associated with it.  
A configuration $\Cfg$ is composed of closed loops, each of which can have a given color and orientation.  For odd $N$, we can also have a color-neutral unoriented loop.  All loop configurations are allowed as long as they do not touch each other.

The weight of a configuration is defined as a product of weights of all the bonds between sites.  Each site can be either empty (Fock vacuum, $|\vac\>_x$), or have two bonds, incoming and outgoing.  For every spatial bond, regardless of color, we pick up a factor of $W_s = e^{-\varepsilon J_1}$ and for every temporal bond, we get $W_t = \varepsilon J_2$.  

The above derivation assumes $\varepsilon$ to be small.  Therefore, to recover the exact results  we can either develop a continuous time formulation \cite{beard_squarelattice_1998a, beard_simulations_1996, beard_study_2005} or we must perform computations at several values of $\varepsilon$ and extrapolate to the $\varepsilon \to 0$ limit.  However, we can also consider the above worldline formulation as the definition of an $O(N)$ model on a space-time lattice, and set the weights of the temporal and spatial hops to be the same.  This gives us a ``relativistic'' model  with manifest symmetry between space and time. Indeed, this relativistic model is sufficient for our purpose and we shall restrict ourselves to it going forward.

\subsection{Worm algorithm for the $O(N)$ qubit model}
\label{sec:qubit-worm}

A major motivation for constructing a worldline representation is that such worldline configurations can be efficiently sampled with local updates using worm algorithms
\cite{prokofev_worm_2010, prokofev_worm_2001}.
In many cases, worm algorithms with a worldline representation also solve the sign problem
\cite{azcoiti_geometric_2009, wolff_simulating_2010, wolff_simulating_2009a, wolff_simulating_2009, wenger_efficient_2009, huffman_fermion_2017, chandrasekharan_fermion_2010}.
In this section, we describe a worm algorithm for the worldline representation constructed above.
Our algorithm is very similar to the one described in Refs.~\cite{singh_qubit_2019, banerjee_conformal_2019}, suitably generalized to $n$ colors.

To design a worm algorithm algorithm, we consider an extended partition function
\begin{align}
  \Ztotal = Z + \Zworm
\end{align}
where $Z$ is the original partition function  from \cref{eq:Z}, and $\Zworm$ is the partition function for the ``worm'' sector, defined as follows. The partition function $\Zworm$ is a sum over configurations similar to $Z$, except that they also contain a creation and annihilation operator insertion at any two space-time points. Therefore the worldline configurations in the worm sector will have several closed worldlines and also one open worldline, with the two ends being the operator insertions. We refer to the open worldline as the ``worm,''  and the two ends of the worm are referred to as the worm ``head'' and ``tail''.

A worm algorithm works by introducing the worm head and tail at a random spacetime site of a configuration in the default sector $Z$.
This starts the worm update, where we sample configurations in the $\Zworm$ sector, by moving the worm head around using local moves satisfying detailed balance.
After each step, we obtain a new configuration in the $\Zworm$ sector.
Finally, when the worm head meets the worm tail, the worm update ends and we end up with a new configuration in the $Z$ sector. 

Our worm algorithm has three types of updates. There is a \emph{begin/end} update, which flips between the $Z$ and the $\Zworm$ sectors. Once inside the worm sector, we have \emph{move} updates, which sample configurations in $\Zworm$ by locally moving the worm head. Finally, we have global \emph{color/orientation flip} updates which change the color or orientation of an entire closed loop at once.
We now describe each of these updates for the worm algorithm.
The updates described below are for the case of oriented worldlines, which implies that there is a unique incoming and outgoing bond at each site. The case of unoriented worldlines can also be handled with minor modifications.

\subsubsection{Begin/end updates}

\begin{figure*}[tbp]
    \centering
    \includegraphics[width=0.8\linewidth]{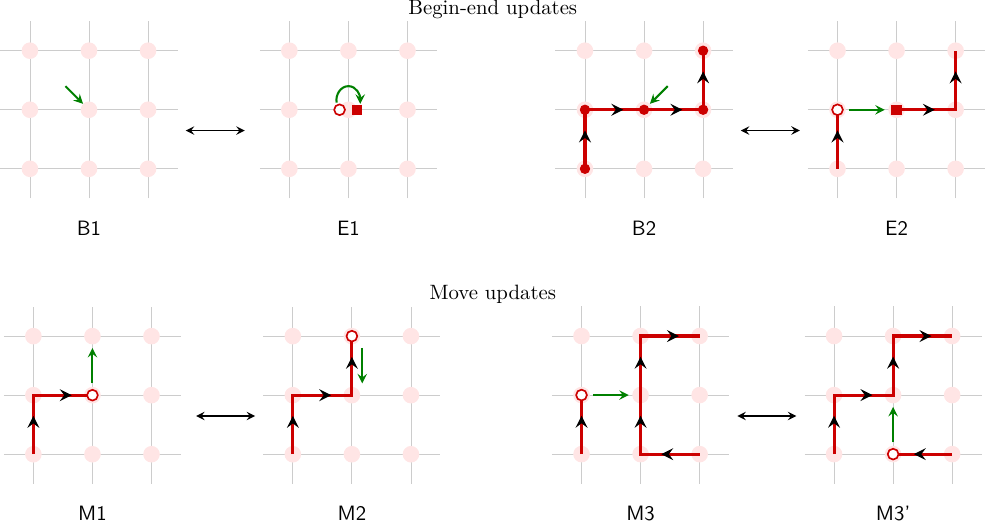}
    \caption[Local configurations for the worm head.]{%
      Local configurations for the worm head.
      We show pairs of configurations that can transform into each other during a worm update with probabilities obeying detailed balance.
      B1 and B2 are not worm configurations, but are the two local configurations where a worm update can begin.
      All other configurations shown are local configurations around the worm head.
      E1, E2 are the two ways a worm update can end.
      The configuration pairs (M$1$, M$2$) and (M$3$,M$3'$) are local configurations that can transform into each other during the ``move'' updates.
    }
    \label{fig:qubit-updates-all}
\end{figure*}

Starting from a configuration in the $Z$ sector, we enter the worm sector using the begin/end update.
We first pick a random spacetime site $x$ to insert the worm head and tail.
We can have two different types of updates depending on the local configuration around the site $x$, as shown in \cref{fig:qubit-updates-all}. Detailed balance implies that for each type of begin update $B$, the reverse update $E$ must also be allowed, which ends the worm update. So, we discuss pairs of configurations $B \leftrightarrow E$, which can transform into each other:

\begin{enumerate}
\item $\text{B}1 \leftrightarrow \text{E}1$: The first possibility is that the site $x$ is empty (local configuration B1 in \cref{fig:qubit-updates-all}).
  In this case, we select a color $c$ at random, and propose to create a worm head and tail of color $c$ at the site $x$.
  The reverse move is that if the worm head and tail are at the same site (local configuration E1), then we propose to exit the worm sector.
\item $\text{B}2 \leftrightarrow \text{E}2$: The second possibility is that the randomly chosen site $x$ is already filled -- that is, it has an incoming and an outgoing bond of some color $c$.
  Let the neighbor in the direction of the incoming bond be $y$.
  In this case, we propose to delete the incoming bond $\<xy\>$, place the worm tail at $x$, and place the worm head at $y$.
  That is, we propose to create the local configuration $\text{E}2$ to enter the worm sector.
  Conversely, the worm update can end at $\text{E}2$, if we propose to move the worm head to a neighboring site which contains the worm tail.
\end{enumerate}

\subsubsection{Move updates}

In the worm sector, we move the worm head around according to \emph{move} updates, shown in bottom row of  \cref{fig:qubit-updates-all}.

For a move update, the first step is to randomly choose  a direction for the worm head to move towards. We do this by choosing any of the $2D$ spacetime directions with equal probability.
Let the neighboring spacetime site in the chosen direction be $y$.
Now, the type of update depends on the local configuration at $y$.
If $y$ contains a worm tail, then the updates E1, E2 apply and can end the worm update, as discussed previously.
However, if $y$ does not contain the worm tail, then the local configuration can be of type M1, M2 or M3. For each of them, we have the following moves:

\begin{itemize}
\item  $\text{M}1 \leftrightarrow  \text{M}2$:
  If the site $y$ is empty (M1), then we simply propose to move the worm head from $x$ to $y$.
  This will take the configuration to M1 $\to $ M2.
  On the other hand, the reverse M2 $\to$ M1 happens if the proposed direction is itself towards an incoming bond. That is, if $\< x y \>$ is the incoming bond at $x$, then we propose to delete the incoming bond $\< x y\>$ and move the worm head to $y$.
\item  $\text{M}3 \leftrightarrow \text{M}3'$:
  The other possibility is that the neighboring site $y$  is completely filled. That is, there is already an incoming and outgoing bond at $y$. In this case, we look at the color $c'$ of the worldline passing through $y$. Let $c$ is the color of the worm head. If $c \neq c'$, then we simply reject this proposal. However, if $c=c'$, then we do the following.
  Let the incoming bond at $y$ be $\<zy\>$, where $z$ is a nearest neighbor of $y$ in the direction of that bond.
  We propose to create the bond $\<xy\>$, delete the bond $\<zy\>$, and move the worm head to $z$.
  The reverse of this move is an identical move.
\end{itemize}

Finally, we note that if a worldline is unoriented, then there is no notion of incoming and outgoing bonds.
For such worldlines, we can use the above updates with small modifications. 
For example, whenever we need to select an incoming or outgoing bond, we instead select one of the two bonds at random, while making sure that detailed balance is satisfied.

\subsubsection{Global color/orientation flip}

Finally, in addition to the above local updates, we add another type of global update. These updates change the color and orientation of an entire loop at once. 
These updates happen in the default sector $Z$.  We start by picking a site at random. If the selected site does not have a worldline passing through it, then we do nothing. However, if the site does have a worldline passing through it, then we propose to randomly change the color and orientation of the worldline. 
Such global updates are very useful since they make it much easier for the algorithm to explore the space of worldline configurations.
We find that this update drastically reduces autocorrelation times, especially for larger $N$.

\subsection{Observables}
\label{sec:qubit-observables}

The worm algorithm allows to compute many observables quite efficiently. For this work, the most important ones are the two-point function susceptibility, and the $O(N)$ current-current susceptibility. However, we measure a couple of other useful observables, which also act as additional non-trivial checks for the code, especially when comparing against exact results from small lattices.

The first, and the simplest, is the \emph{vacuum density}, which we define as the average density of singlet sites 
\begin{align}
v &= \frac{1}{Z} \Tr\left( \frac{1}{L^d} \sum_{x} P^{\vac}_{x} e^{-\beta H} \right),
\end{align}
where $P^{\vac}_{x} = |\vac \>_x \< \vac |_x$ is the projector onto the singlet state at the site $x$, and $d$ is the number spatial dimensions.
For a given worldline configuration, this can be measured simply by counting the total number of vacuum sites on the spacetime lattice. 

Next, we measure the $n$ types of $O(N)$ charges (defined earlier in \cref{sec:qubit-model}), given by
\begin{align}
\langle Q_{k} \rangle = \frac{1}{Z} \Tr\left( \sum_{x} \hat Q_{x, k} e^{-\beta H} \right)\quad \text{for } k=1,\dotsc,n\ ,
\end{align}
where the operator $\hat Q_{x,k}$ measures the local charge  $\vec q^{\pm k}$  at the site $x$,
\begin{align}
  \hat Q_{k,x} = c^{\dagger}_{x,k} c_{x,k}^{\vphantom{\dagger}} -  c^{\dagger}_{x,-k} c_{x,-k}^{\vphantom{\dagger}}\quad \text{for } k=1,\dotsc,n.
\end{align}
In our formulation, the worldlines have well-defined charges by construction. Therefore, we  can measure the $O(N)$ charges by picking a spatial slice for each configuration in the default sector $Z$, and simply counting the number of worldlines passing through it having some given color $k$. 
Note that for the cases considered in this work, the expectation value of the $O(N)$ charges are zero since the partition function has an $O(N)$ symmetry.
However, it can still be useful to measure this observable 
because we can study sectors of fixed non-zero global charge by introducing a chemical potential for the charges.

An important observable for us is the $O(N)$ current-current susceptibility, which can be computed by measuring the winding number,
\begin{align}
  \rho_s^{(k)} = \frac{1}{L^{d-2} \beta} \langle (N_{w}^{(k)})^2 \rangle, \quad \text{for } k=1,\dotsc,n,
  \label{eq:spin-stiffness-defn}
\end{align}
where $d$ is the number of spatial dimensions,
$k=1,\dotsc,n$ is a given color,  and $N_w^{(k)}$ is the winding number of color-$k$ worldlines in a configuration. The winding number $N_w^{(k)}$ is measured by picking a plane $(x,t)$, where $x$ can be any spatial direction and $t$ is the temporal direction, and then counting the number of worldlines of a given color $k$ passing through it.
In our computations, the $O(N)$ symmetry ensures that $\rho_s^{(k)}$ will be identical for all colors $k$.
Therefore, we average over all colors $k$ to improve the statistics.
From now on, we drop the color superscript $k$, and just denote this color-averaged observable by $\rho_s$.

The final observable is the susceptibility of the two-point correlation function of charge $\vec q^{\pm k}$ particles,
\begin{align}
  \chi_{k} = \frac{1}{Z \beta L^{2d} } \sum_{x,y} &\int_0^\beta\!\! dt\, \Tr\left( {e}^{-(\beta - t) H} \hat O_{y, k} {e}^{-t H} \hat O^\dagger_{x,k}\right),
\label{eq:suscep-defn}
\end{align}
for $k=0,1\dotsc,n$ (omitting $k=0$ if $N$ is even), and the operator $\hat O_{x,k}$, defined as
\begin{align}
  \hat O^\dagger_{x,k} = (c^\dagger_{x,k} + c_{x, -k }),
\end{align}
creates particles and annihilates anti-particles of charge $\vec q^k$.
The worm algorithm already works as an improved estimator for this observable, since the worm sector samples configurations with $O_{x,k}$ operator insertions. Therefore, to compute $\chi_k$, we just have to count the number of configurations generated in a given worm update. The expectation value of this number is exactly $\chi_k$, up to normalization.
As before, we average over all colors to improve statistics, and denote the color-averaged susceptibility as $\chi$.

\begin{figure*}[htp]
    \centering
    \includegraphics[width=0.49\textwidth]{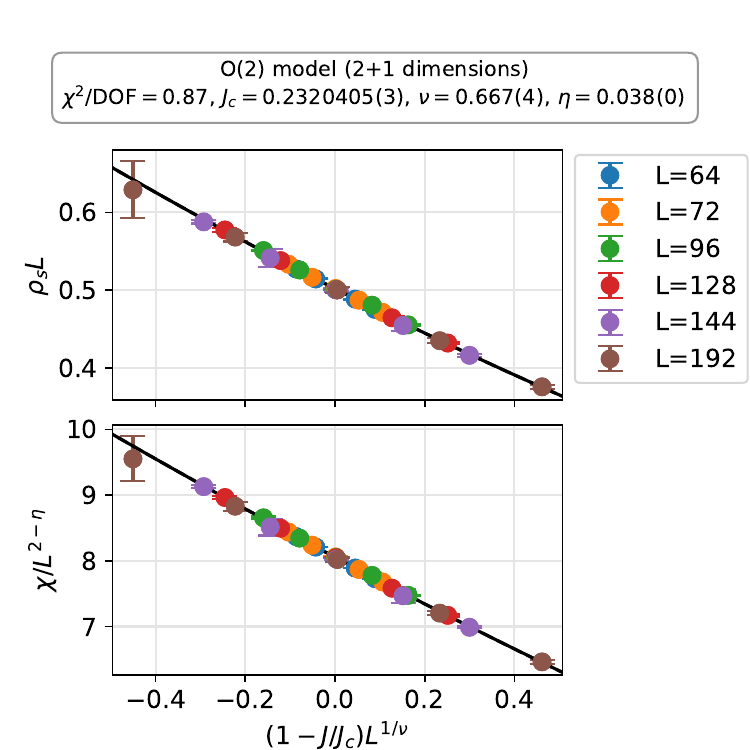}
    \includegraphics[width=0.49\textwidth]{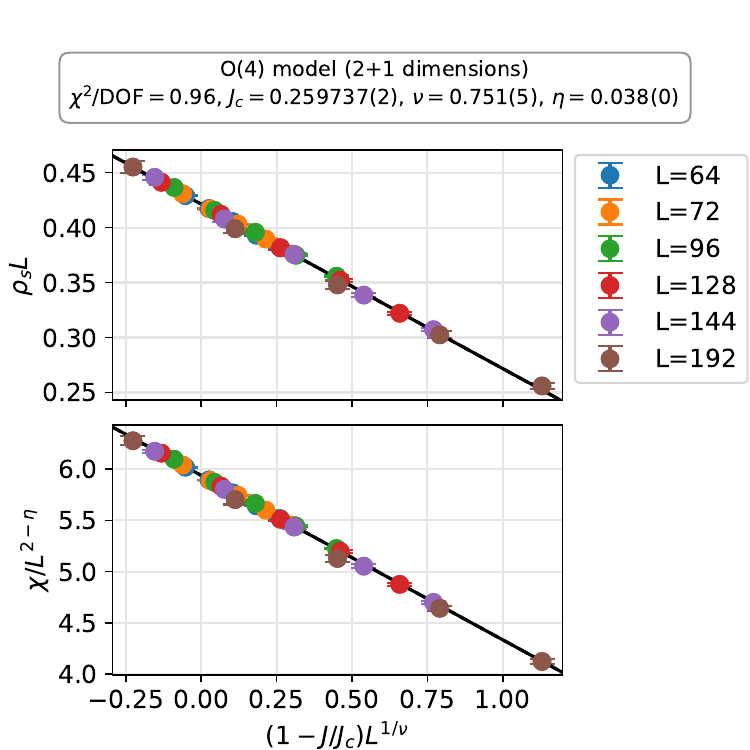}\\
    \includegraphics[width=0.49\textwidth]{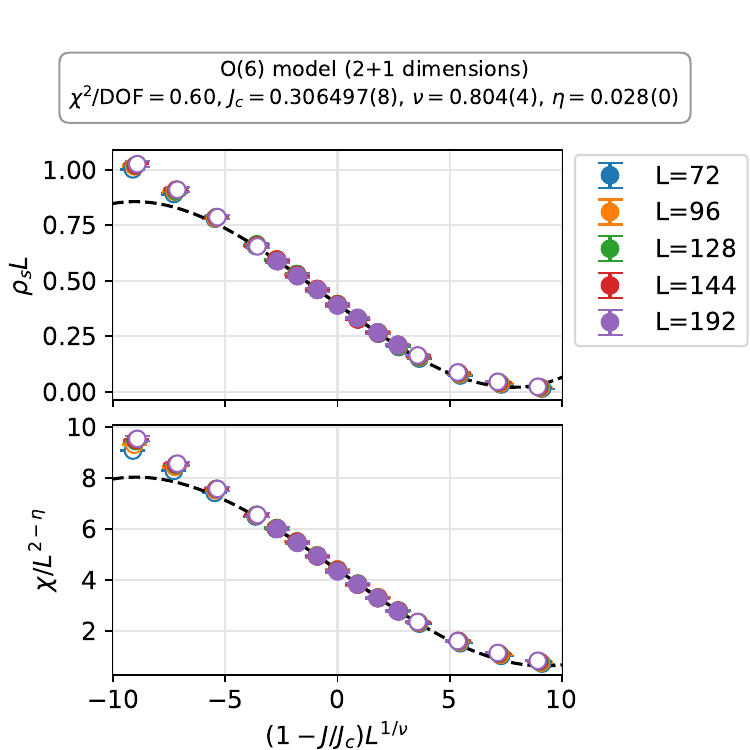}
    \includegraphics[width=0.49\textwidth]{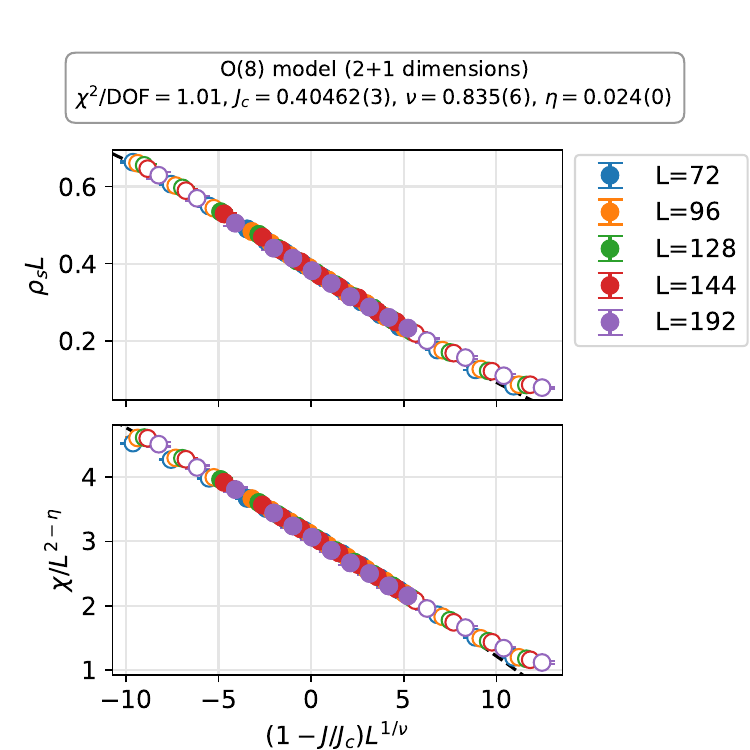}
    \caption{
      Scaling collapse for the $O(N)$ qubit models close to the critical point, for $N=2,4,6,8$.
      The top plot for each $N$ shows the data for winding-number susceptibility $\rho_s$ while the bottom plot shows the two--point function susceptibility $\chi$.
      The circles show Monte Carlo data for various $L$.
      The black line shows a combined fit to both $\rho_s(J,L)$ and $\chi(J,L)$ for various $J,L$.
      The filled circles were used in the fit as the scaling window, while the open circles were not used in the fit.
      We find that even the open circles all lie on the same curve, within small deviations expected at the edges from corrections to scaling, indicating that the scaling hypothesis is well satisfied in the fitting regime.
      In the final fits, the values of the critical exponent $\eta$ as well as the $y$-intercepts $f(0)$, $g(0)$ were fixed by performing the fits shown in \cref{fig:critical-scaling-1}.
      For more details on the fitting procedure, see \cref{sec:fits-scaling}.
    }
    \label{fig:critical-scaling-1}
\end{figure*}

\begin{figure*}[ht]
    \centering
    \includegraphics[width=0.49\linewidth]{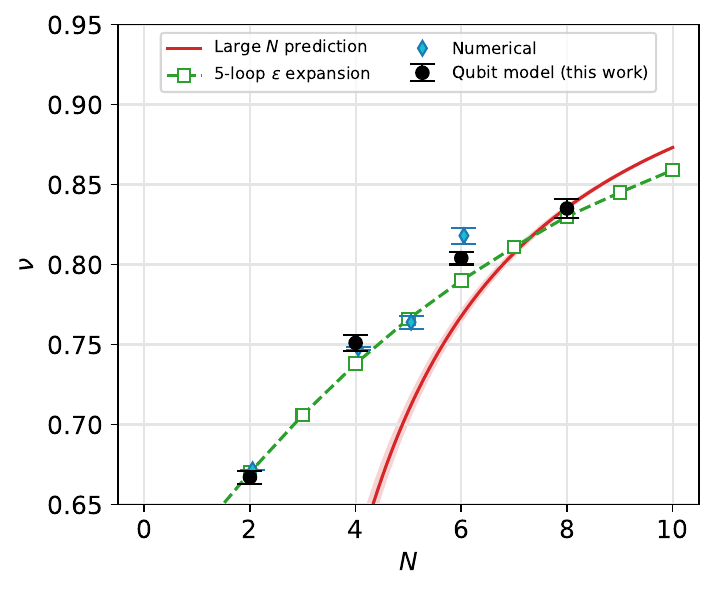}\hfill
    \includegraphics[width=0.49\linewidth]{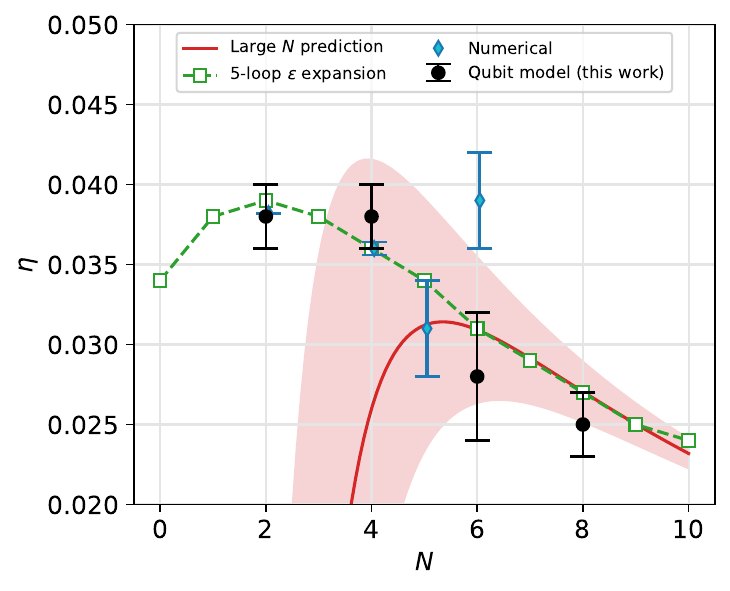}
    \caption{%
      Comparison of the qubit $O(N)$ model against existing results in literature for the \ac{WF} critical exponents $\nu$, $\eta$.
      The black dots labeled ``qubit model'' are from this work.
      The green boxes for the $\varepsilon$-expansion are from Ref.~\cite{antonenko_critical_1995}.
      The large-$N$ results \cite{moshe_quantum_2003} are shown as a solid red line,
      with the shaded band showing expected higher order $1/N$ contributions assuming they are of natural size.
      We show the best numerical results from lattice Monte Carlo and conformal bootstrap as blue diamonds.
      The numbers for the critical exponents, along with the references, are listed in \cref{tab:comparison}.
      We find excellent agreement with all the other techniques.}
    \label{fig:large-n-comparison}
\end{figure*}

\section{Results: Wilson-Fisher fixed point}
\label{sec:criticality}

Using the worm algorithm described in the previous section, we have performed computations for the $O(N)$ models for various $N$.
To show that the physics of the continuum $O(N)$ \ac{NLSM} is reproduced in the qubit models, we locate the critical point and compute the critical exponents $\nu$ and $\eta$. 
We perform computations for $N=2,4,6,8$ and lattice sizes up to $L=192$.
While the model is well-defined for any $N$, it seems to show some unexpected behavior for $N \geq 10$, which prevented us from pushing the calculations to higher $N$. We will comment on this later in the conclusions.

\sisetup{table-number-alignment = center, table-format = 2.8, table-align-text-post = false }
\ctable[
caption  = Comparison of the qubit model results with literature,
label    = tab:comparison,
star, center, doinside = \smaller]%
{l S S S S @{\extracolsep{1em}} S S S S @{\extracolsep{0.1em}} S}%
{%
\tnote[a]{This work}
\tnote[b]{Conformal bootstrap \cite{chester_carving_2020}}
\tnote[c]{Lattice Monte Carlo \cite{deng_bulk_2006}}
\tnote[d]{Lattice Monte Carlo \cite{butti_critical_2005}}
\tnote[e]{Lattice Monte Carlo \cite{holtmann_critical_2003}}

}{\FL
   & \multicolumn{4}{c}{$\nu$} & \multicolumn{4}{c}{$\eta$} & \NN
   \cmidrule{2-5}
   \cmidrule{6-9}
  N                    & 2                     & 4                    & 6                   & 8          & 2                     & 4                    & 6                   & 8          & \NN
  Qubit model\tmark[a] & 0.6670(40)            & 0.7510(50)           & 0.8040(40)          & 0.8350(60) & 0.0380(20)            & 0.0380(20)           & 0.0280(40)          & 0.0250(20) & \NN 
  $\epsilon$ expansion        & 0.670                 & 0.738                & 0.790               & 0.830      & 0.039                 & 0.036                & 0.031               & 0.027      & \NN
  Large $N$            & -0.010                & 0.612                & 0.768               & 0.836      & -0.149                & 0.026                & 0.031               & 0.027      & \NN
Numerical              & 0.671754(81)\tmark[b] & 0.74766(84)\tmark[c] & 0.8180(50)\tmark[d] &            & 0.038176(34)\tmark[b] & 0.03600(40)\tmark[c] & 0.0390(30)\tmark[d] &            & \LL
}

Near a critical point $J=J_c$, the two observables $\rho_s$ and susceptibility $\chi$ should have the scaling behavior
\begin{align}
    \rho_s(u; L) &= \frac{1}{L^{d-2}} f(u) \\
    \chi(u; L) &= {L^{2-\eta}} g(u),
              \label{eq:scaling}
\end{align}
where $u=(J-J_c) L^{1/\nu}$ is the scaling variable, $f(u)$ and $g(u)$ are unknown universal functions, and $\eta$ and $\nu$ are the critical exponents.  We can approximate $f(u)$ and $g(u)$ by polynomials and perform a combined fit to $\rho_s$ and $\chi$, which helps us extract $J_c$, $\nu$, $\eta$, and the functions $f(u)$ and $g(u)$.  
To gain better precision and control over the fits, we perform a few additional steps. Our fitting procedure is described in detail in \cref{sec:fits-scaling}.
The results of these fits to the scaling behavior in \cref{eq:scaling} are shown in \cref{fig:critical-scaling-1}. As can be seen from the figure, we find excellent scaling collapse, which lets us obtain the location of the critical point and the critical exponents.

To establish the universality class, we perform a comparison of our computations with existing results in the literature.
The critical properties of the $2+1d$ $O(N)$ \ac{NLSM} have been studied extensively through several numerical and analytic techniques.
Numerically, precise results are available from lattice Monte Carlo \cite{hasenbusch_anisotropic_2011, campostrini_critical_2006, hasenbusch_monte_2019, pelissetto_critical_2002} and conformal bootstrap techniques \cite{chester_carving_2020, kos_precision_2016}.
On the analytic side, even though the $2+1d$ $O(N)$ \ac{NLSM} is a strongly interacting field theory, several techniques have been developed to leverage a ``hidden'' small parameter and enable analytic control, such as the $1/N$ and $\varepsilon$ expansions.
We summarize our results and show a comparison with other results from literature in \cref{fig:large-n-comparison} and \cref{tab:comparison}.

First, we consider the critical exponents for the $O(N)$ model from the large-$N$ expansion, which is exact in the $N \to \infty$ limit.
In the $1/N$ expansion, the critical exponent $\nu$ is known up to order $N^{-2}$ and $\eta$ up to order $N^{-3}$ \cite{moshe_quantum_2003},
\begin{align}
  \nu &= 1-\frac{32}{3 \pi ^2} \frac1N  + \left(\frac{3584}{27 \pi ^4}-\frac{32}{\pi ^2}\right) \frac1{N^{2}} +O(N^{-3}), \\
  \eta &= \frac{\eta_1}{N} + \frac{\eta_2}{N^2} + \frac{\eta_3}{N^3} + O(N^{-4}),
    \label{eq:large-n-results}
\end{align}
with
\begin{align}
  \eta_1 &= \frac{8}{3\pi^2},\quad \\
  \eta_2 &=-\frac83 \eta_1^2, \quad  \\
  \eta_3 &= \eta_1^3 \left[ -\frac{797}{18} - \frac{61}{24} \pi^2 + \frac{27}{8} \psi''(1/2)  + \frac92 \pi^2 \ln 2\right],
\end{align}
where $\psi(x) \equiv  (d/d \ln x) \Gamma(x)$ is the logarithmic derivative of the $\Gamma$ function.
We show the large-$N$ results as a solid line in \cref{fig:large-n-comparison}.

The next analytic technique we compare with is the $\varepsilon$ expansion.
Since the $\varepsilon$ expansion is a divergent series, several resummation methods have been employed in the literature. However, such resummation techniques are uncontrolled and do not yield precise error estimates.
Hence, we do not show error estimates for the $\varepsilon$-expansion results in \cref{fig:large-n-comparison}.
The values shown are from Ref.~\cite{antonenko_critical_1995}, where the authors computed the critical exponents from a five-loop $\varepsilon$ expansion and tabulated the results up to $N=28$.

Finally, we show the best numerical results known to us for various $N$.
For $N=2$, conformal bootstrap yields the best estimates \cite{kos_precision_2016}, while higher $N \geq 4$ results are from lattice Monte Carlo computations \cite{pelissetto_critical_2002}. In all of these cases, the state-of-the-art results, while more precise, are in agreement with our work.
It is encouraging that we are already able to achieve reasonably good precision with the qubit model and a simple finite-size-scaling analysis. This suggests that, for the qubit models, a precision comparable to the best known results should be within reach using a more sophisticated finite-size scaling analysis.

We consistently find that the critical exponent $\nu$ from the qubit model agrees very well with all the existing results up to $N=8$.
The critical exponent $\eta$, being much smaller, has larger error bars on this scale in \cref{fig:large-n-comparison}.
It is harder to extract $\eta$ to a high relative precision without a more sophisticated finite-size scaling analysis, which could be the subject of a future study.
We do find that the results from other techniques for $\eta$ lie within $1\sigma$ of our computations. Based on all these comparisons, we conclude that the qubit regularized $O(N)$ models constructed in this work indeed have a second-order critical point that lies in the $O(N)$ \ac{WF} universality class, at least up to $N=8$.

\section{Discussion and conclusions}
\label{sec:conclusions}

In this work, we developed a qubit regularization of the $O(N)$ \ac{NLSM} for $D \geq 3 $ spacetime dimensions.  We provided strong numerical evidence that this model has a critical point in the $O(N)$ \ac{WF} universality class.
In particular, we computed the critical exponents $\nu, \eta$ for  $N = 2,4,6,8$, and compared these with the available results in the literature for the \ac{WF} fixed point from various other techniques.

A major motivation of this work was to develop the idea of qubit regularization of \acp{QFT}, proposed in Ref.~\cite{singh_qubit_2019}.
There, the authors constructed a qubit model for the $O(3)$ \ac{NLSM} using two qubits per lattice site.
In this work, we demonstrate the two-qubit $O(3)$ model of Ref.~\cite{singh_qubit_2019} can be generalized to an $O(N)$ model for arbitrary $N$, although with a slightly larger Hilbert space ($N+1$ dimensional) at each lattice site.

A particularly appealing feature of such qubit models is their suitability for implementation on quantum computers.
When the quantum critical point can be located precisely, say using classical \ac{MC} computations, then a continuum \ac{QFT} is guaranteed to emerge. Close to this critical point, we may study any number of dynamical observables on a quantum computer, which would otherwise be typically inaccessible on classical computers. At least in the \textsc{nisq}-era, it will be extremely important to construct models that use the available qubits economically, and yet reproduce the physics of a \ac{QFT} to desired precision.

While we had hoped to push our calculations for $N \geq 10$ as well, we encountered a somewhat puzzling phenomenon.
We found that the higher $N$ qubit models require larger volumes, which are limited by available computational resources.
This suggests that there may be a new scale in the problem that is controlled by $N$.
It also looks like, up to the lattice sizes considered in this work, the $N=10$ qubit model is not in the usual superfluid phase for large negative $\lambda$, as it is in the $N\leq8$ models.
Whether this signals a new kind of phase transition, or is simply an artifact of smaller volumes, is not clear.
It would be very interesting to systematically explore higher $N$ qubit models. We leave this for a follow-up work.

Recently, Ref.~\cite{liu_qubit_2022} studied the $O(N)$ qubit models from an algebraic perspective. They made the interesting observation that the qubit model algebra for the $O(N)$ model with the singlet and fundamental representations (like the one considered here) reproduces the infinite-dimensional $O(N)$ algebra in the large-$N$ limit. Naively, this might suggest that the large-$N$ qubit models should converge faster to the continuum limit, which seems to be in some tension with the 
large-$N$ behavior of our models.
Perhaps there might be other large-$N$ qubit models which reproduce the continuum physics faster, in agreement with the analysis of the qubit algebras. This would be interesting to clarify and explore in a future work.

All of this is especially interesting in two spacetime dimensions, which is a case we did not explore in this work.  This is because the continuum $O(N)$ models are known to be asymptotically free in two spacetime dimensions for $N\geq 3$.
The $O(3)$ model also allows for a topological $\theta$ term, and has been long studied as a toy model of \ac{QCD}.
While it has been long known that the continuum nonlinear sigma models can be obtained from the spin-$S$ antiferromagnetic Heisenberg chain
\cite{affleck_critical_1987a, haldane_continuum_1983, haldane_nonlinear_1983}
in the large-$S$ limit, or via dimensional-reduction of a $2+1d$ Heisenberg antiferromagnetic within a D-theory approach
\cite{chandrasekharan_quantum_1997, brower_qcd_1999, chandrasekharan_spin_2002, brower_dtheory_2004, chandrasekharan_spin_2002}, it was not clear whether the continuum limit of such asymptotically free theories can be reached with a fixed \emph{finite} local Hilbert space.
Indeed, in our investigations, the $O(N)$ qubit models considered in this work do not have a second-order critical point in two-dimensions in the usual limit, as has also been noted in several other works \cite{bruckmann_nonlinear_2019, niedermayer_model_2016, zhou_spacetime_2021}.
Interestingly, Ref.~\cite{bhattacharya_qubit_2021} recently found that a very simple $SU(2)$-symmetric two-qubit model, called the Heisenberg comb, indeed displays signatures of asymptotic freedom.
In the context of our work, this raises the exciting possibility of obtaining all other $O(N)$ models, which are also known to asymptotically free for $N\geq 3$, using such simple qubit Hamiltonians.
It would be interesting to explore the $O(N)$-invariant qubit model space in $D=2$ spacetime dimensions further to find out whether the asymptotically free $O(N)$ critical point can be identified.

In another recent related work \cite{zhou_spacetime_2021}, the authors showed that an $O(4)$ model, similar to this work, does in fact display asymptotic freedom, if considered within a D-theory approach by introducing a small additional dimension.  Correlation length diverges exponentially as the length of additional dimension is increased, and the system approaches the $1+1$d continuum $O(4)$ \ac{NLSM}. Based on those results, it is likely that the $O(N)$ models considered in this work also show asymptotic freedom for all $N \geq 3$ in $1+1$ dimensions within a D-theory formulation. This would be interesting to confirm by numerical computations as well, especially with regards to the question how quickly the continuum limit is approached for large $N$.

\section*{Acknowledgments}

I would especially like to thank Shailesh Chandrasekharan and Roxanne Springer for critical feedback.
I would also like to like to thank Alex Buser, Stephan Caspar, Andrew Gasbarro, Domenico Orlando, Hanqing Liu, Mendel Nguyen, Rajan Gupta, Rolando Somma, Ronen Plesser, Susanne Reffert, Tanmoy Bhattacharya and Uwe-Jens Wiese for useful conversations.  

The material presented here was funded 
in part by the DOE QuantISED program through the theory  consortium ``Intersections of QIS and Theoretical Particle Physics'' at Fermilab with Fermilab Subcontract No. 666484,
in part by Institute for Nuclear Theory with US Department of Energy Grant DE-FG02-00ER41132,
in part by U.S. Department of Energy, Office of Science, Office of Nuclear Physics, Inqubator for Quantum Simulation (IQuS) under Award Number DOE (NP) Award DE-SC0020970, 
in part by U.S. Department of Energy, Office of Science, Nuclear Physics program under award No.~DE-FG02-05ER41368, 
and in part by the J.\,Horst Meyer Endowment fellowship from Duke University.

\bibliographystyle{apsrev4-2}
\bibliography{refs}

\begin{thebibliography}{71}%
\makeatletter
\providecommand \@ifxundefined [1]{%
 \@ifx{#1\undefined}
}%
\providecommand \@ifnum [1]{%
 \ifnum #1\expandafter \@firstoftwo
 \else \expandafter \@secondoftwo
 \fi
}%
\providecommand \@ifx [1]{%
 \ifx #1\expandafter \@firstoftwo
 \else \expandafter \@secondoftwo
 \fi
}%
\providecommand \natexlab [1]{#1}%
\providecommand \enquote  [1]{``#1''}%
\providecommand \bibnamefont  [1]{#1}%
\providecommand \bibfnamefont [1]{#1}%
\providecommand \citenamefont [1]{#1}%
\providecommand \href@noop [0]{\@secondoftwo}%
\providecommand \href [0]{\begingroup \@sanitize@url \@href}%
\providecommand \@href[1]{\@@startlink{#1}\@@href}%
\providecommand \@@href[1]{\endgroup#1\@@endlink}%
\providecommand \@sanitize@url [0]{\catcode `\\12\catcode `\$12\catcode
  `\&12\catcode `\#12\catcode `\^12\catcode `\_12\catcode `\%12\relax}%
\providecommand \@@startlink[1]{}%
\providecommand \@@endlink[0]{}%
\providecommand \url  [0]{\begingroup\@sanitize@url \@url }%
\providecommand \@url [1]{\endgroup\@href {#1}{\urlprefix }}%
\providecommand \urlprefix  [0]{URL }%
\providecommand \Eprint [0]{\href }%
\providecommand \doibase [0]{https://doi.org/}%
\providecommand \selectlanguage [0]{\@gobble}%
\providecommand \bibinfo  [0]{\@secondoftwo}%
\providecommand \bibfield  [0]{\@secondoftwo}%
\providecommand \translation [1]{[#1]}%
\providecommand \BibitemOpen [0]{}%
\providecommand \bibitemStop [0]{}%
\providecommand \bibitemNoStop [0]{.\EOS\space}%
\providecommand \EOS [0]{\spacefactor3000\relax}%
\providecommand \BibitemShut  [1]{\csname bibitem#1\endcsname}%
\let\auto@bib@innerbib\@empty
\bibitem [{\citenamefont {Feynman}(1982)}]{feynman_simulating_1982}%
  \BibitemOpen
  \bibfield  {author} {\bibinfo {author} {\bibfnamefont {R.~P.}\ \bibnamefont
  {Feynman}},\ }\href {https://doi.org/10.1007/BF02650179} {\bibfield
  {journal} {\bibinfo  {journal} {International Journal of Theoretical
  Physics}\ }\textbf {\bibinfo {volume} {21}},\ \bibinfo {pages} {467}
  (\bibinfo {year} {1982})}\BibitemShut {NoStop}%
\bibitem [{\citenamefont {Benioff}(1980)}]{benioff_computer_1980}%
  \BibitemOpen
  \bibfield  {author} {\bibinfo {author} {\bibfnamefont {P.}~\bibnamefont
  {Benioff}},\ }\href {https://doi.org/10.1007/BF01011339} {\bibfield
  {journal} {\bibinfo  {journal} {Journal of Statistical Physics}\ }\textbf
  {\bibinfo {volume} {22}},\ \bibinfo {pages} {563} (\bibinfo {year}
  {1980})}\BibitemShut {NoStop}%
\bibitem [{\citenamefont
  {Preskill}(2018{\natexlab{a}})}]{preskill_simulating_2018}%
  \BibitemOpen
  \bibfield  {author} {\bibinfo {author} {\bibfnamefont {J.}~\bibnamefont
  {Preskill}},\ }\href@noop {} {\bibfield  {journal} {\bibinfo  {journal}
  {arXiv:1811.10085 [hep-lat, physics:hep-th, physics:quant-ph]}\ } (\bibinfo
  {year} {2018}{\natexlab{a}})}\BibitemShut {NoStop}%
\bibitem [{\citenamefont
  {Preskill}(2018{\natexlab{b}})}]{preskill_quantum_2018}%
  \BibitemOpen
  \bibfield  {author} {\bibinfo {author} {\bibfnamefont {J.}~\bibnamefont
  {Preskill}},\ }\href {https://doi.org/10.22331/q-2018-08-06-79} {\bibfield
  {journal} {\bibinfo  {journal} {Quantum}\ }\textbf {\bibinfo {volume} {2}},\
  \bibinfo {pages} {79} (\bibinfo {year} {2018}{\natexlab{b}})}\BibitemShut
  {NoStop}%
\bibitem [{\citenamefont {Dalmonte}\ and\ \citenamefont
  {Montangero}(2016)}]{dalmonte_lattice_2016}%
  \BibitemOpen
  \bibfield  {author} {\bibinfo {author} {\bibfnamefont {M.}~\bibnamefont
  {Dalmonte}}\ and\ \bibinfo {author} {\bibfnamefont {S.}~\bibnamefont
  {Montangero}},\ }\href {https://doi.org/10.1080/00107514.2016.1151199}
  {\bibfield  {journal} {\bibinfo  {journal} {Contemporary Physics}\ }\textbf
  {\bibinfo {volume} {57}},\ \bibinfo {pages} {388} (\bibinfo {year}
  {2016})}\BibitemShut {NoStop}%
\bibitem [{\citenamefont {Ba{\~n}uls}\ \emph {et~al.}(2019)\citenamefont
  {Ba{\~n}uls}, \citenamefont {Blatt}, \citenamefont {Catani}, \citenamefont
  {Celi}, \citenamefont {Cirac}, \citenamefont {Dalmonte}, \citenamefont
  {Fallani}, \citenamefont {Jansen}, \citenamefont {Lewenstein}, \citenamefont
  {Montangero}, \citenamefont {Muschik}, \citenamefont {Reznik}, \citenamefont
  {Rico}, \citenamefont {Tagliacozzo}, \citenamefont {Van~Acoleyen},
  \citenamefont {Verstraete}, \citenamefont {Wiese}, \citenamefont {Wingate},
  \citenamefont {Zakrzewski},\ and\ \citenamefont
  {Zoller}}]{banuls_simulating_2019}%
  \BibitemOpen
  \bibfield  {author} {\bibinfo {author} {\bibfnamefont {M.~C.}\ \bibnamefont
  {Ba{\~n}uls}}, \bibinfo {author} {\bibfnamefont {R.}~\bibnamefont {Blatt}},
  \bibinfo {author} {\bibfnamefont {J.}~\bibnamefont {Catani}}, \bibinfo
  {author} {\bibfnamefont {A.}~\bibnamefont {Celi}}, \bibinfo {author}
  {\bibfnamefont {J.~I.}\ \bibnamefont {Cirac}}, \bibinfo {author}
  {\bibfnamefont {M.}~\bibnamefont {Dalmonte}}, \bibinfo {author}
  {\bibfnamefont {L.}~\bibnamefont {Fallani}}, \bibinfo {author} {\bibfnamefont
  {K.}~\bibnamefont {Jansen}}, \bibinfo {author} {\bibfnamefont
  {M.}~\bibnamefont {Lewenstein}}, \bibinfo {author} {\bibfnamefont
  {S.}~\bibnamefont {Montangero}}, \bibinfo {author} {\bibfnamefont {C.~A.}\
  \bibnamefont {Muschik}}, \bibinfo {author} {\bibfnamefont {B.}~\bibnamefont
  {Reznik}}, \bibinfo {author} {\bibfnamefont {E.}~\bibnamefont {Rico}},
  \bibinfo {author} {\bibfnamefont {L.}~\bibnamefont {Tagliacozzo}}, \bibinfo
  {author} {\bibfnamefont {K.}~\bibnamefont {Van~Acoleyen}}, \bibinfo {author}
  {\bibfnamefont {F.}~\bibnamefont {Verstraete}}, \bibinfo {author}
  {\bibfnamefont {U.-J.}\ \bibnamefont {Wiese}}, \bibinfo {author}
  {\bibfnamefont {M.}~\bibnamefont {Wingate}}, \bibinfo {author} {\bibfnamefont
  {J.}~\bibnamefont {Zakrzewski}},\ and\ \bibinfo {author} {\bibfnamefont
  {P.}~\bibnamefont {Zoller}},\ }\href@noop {} {\bibfield  {journal} {\bibinfo
  {journal} {arXiv:1911.00003 [cond-mat, physics:hep-lat, physics:hep-th,
  physics:quant-ph]}\ } (\bibinfo {year} {2019})}\BibitemShut {NoStop}%
\bibitem [{\citenamefont {Ba{\~n}uls}\ and\ \citenamefont
  {Cichy}(2019)}]{banuls_review_2019}%
  \BibitemOpen
  \bibfield  {author} {\bibinfo {author} {\bibfnamefont {M.~C.}\ \bibnamefont
  {Ba{\~n}uls}}\ and\ \bibinfo {author} {\bibfnamefont {K.}~\bibnamefont
  {Cichy}},\ }\href@noop {} {\bibfield  {journal} {\bibinfo  {journal}
  {arXiv:1910.00257 [hep-lat, physics:hep-th, physics:quant-ph]}\ } (\bibinfo
  {year} {2019})}\BibitemShut {NoStop}%
\bibitem [{\citenamefont {Alexeev}\ \emph {et~al.}(2020)\citenamefont
  {Alexeev}, \citenamefont {Bacon}, \citenamefont {Brown}, \citenamefont
  {Calderbank}, \citenamefont {Carr}, \citenamefont {Chong}, \citenamefont
  {DeMarco}, \citenamefont {Englund}, \citenamefont {Farhi}, \citenamefont
  {Fefferman}, \citenamefont {Gorshkov}, \citenamefont {Houck}, \citenamefont
  {Kim}, \citenamefont {Kimmel}, \citenamefont {Lange}, \citenamefont {Lloyd},
  \citenamefont {Lukin}, \citenamefont {Maslov}, \citenamefont {Maunz},
  \citenamefont {Monroe}, \citenamefont {Preskill}, \citenamefont {Roetteler},
  \citenamefont {Savage},\ and\ \citenamefont
  {Thompson}}]{alexeev_quantum_2020}%
  \BibitemOpen
  \bibfield  {author} {\bibinfo {author} {\bibfnamefont {Y.}~\bibnamefont
  {Alexeev}}, \bibinfo {author} {\bibfnamefont {D.}~\bibnamefont {Bacon}},
  \bibinfo {author} {\bibfnamefont {K.~R.}\ \bibnamefont {Brown}}, \bibinfo
  {author} {\bibfnamefont {R.}~\bibnamefont {Calderbank}}, \bibinfo {author}
  {\bibfnamefont {L.~D.}\ \bibnamefont {Carr}}, \bibinfo {author}
  {\bibfnamefont {F.~T.}\ \bibnamefont {Chong}}, \bibinfo {author}
  {\bibfnamefont {B.}~\bibnamefont {DeMarco}}, \bibinfo {author} {\bibfnamefont
  {D.}~\bibnamefont {Englund}}, \bibinfo {author} {\bibfnamefont
  {E.}~\bibnamefont {Farhi}}, \bibinfo {author} {\bibfnamefont
  {B.}~\bibnamefont {Fefferman}}, \bibinfo {author} {\bibfnamefont {A.~V.}\
  \bibnamefont {Gorshkov}}, \bibinfo {author} {\bibfnamefont {A.}~\bibnamefont
  {Houck}}, \bibinfo {author} {\bibfnamefont {J.}~\bibnamefont {Kim}}, \bibinfo
  {author} {\bibfnamefont {S.}~\bibnamefont {Kimmel}}, \bibinfo {author}
  {\bibfnamefont {M.}~\bibnamefont {Lange}}, \bibinfo {author} {\bibfnamefont
  {S.}~\bibnamefont {Lloyd}}, \bibinfo {author} {\bibfnamefont {M.~D.}\
  \bibnamefont {Lukin}}, \bibinfo {author} {\bibfnamefont {D.}~\bibnamefont
  {Maslov}}, \bibinfo {author} {\bibfnamefont {P.}~\bibnamefont {Maunz}},
  \bibinfo {author} {\bibfnamefont {C.}~\bibnamefont {Monroe}}, \bibinfo
  {author} {\bibfnamefont {J.}~\bibnamefont {Preskill}}, \bibinfo {author}
  {\bibfnamefont {M.}~\bibnamefont {Roetteler}}, \bibinfo {author}
  {\bibfnamefont {M.}~\bibnamefont {Savage}},\ and\ \bibinfo {author}
  {\bibfnamefont {J.}~\bibnamefont {Thompson}},\ }\href@noop {} {\bibfield
  {journal} {\bibinfo  {journal} {arXiv:1912.07577 [cond-mat, physics:hep-lat,
  physics:hep-th, physics:nucl-th, physics:quant-ph]}\ } (\bibinfo {year}
  {2020})}\BibitemShut {NoStop}%
\bibitem [{\citenamefont {Kogut}(1979)}]{kogut_introduction_1979}%
  \BibitemOpen
  \bibfield  {author} {\bibinfo {author} {\bibfnamefont {J.~B.}\ \bibnamefont
  {Kogut}},\ }\href {https://doi.org/10.1103/RevModPhys.51.659} {\bibfield
  {journal} {\bibinfo  {journal} {Reviews of Modern Physics}\ }\textbf
  {\bibinfo {volume} {51}},\ \bibinfo {pages} {659} (\bibinfo {year}
  {1979})}\BibitemShut {NoStop}%
\bibitem [{\citenamefont {Wilson}(1983)}]{wilson_renormalization_1983}%
  \BibitemOpen
  \bibfield  {author} {\bibinfo {author} {\bibfnamefont {K.~G.}\ \bibnamefont
  {Wilson}},\ }\href {https://doi.org/10.1103/RevModPhys.55.583} {\bibfield
  {journal} {\bibinfo  {journal} {Reviews of Modern Physics}\ }\textbf
  {\bibinfo {volume} {55}},\ \bibinfo {pages} {583} (\bibinfo {year}
  {1983})}\BibitemShut {NoStop}%
\bibitem [{\citenamefont {Wilson}\ and\ \citenamefont
  {Kogut}(1974)}]{wilson_renormalization_1974}%
  \BibitemOpen
  \bibfield  {author} {\bibinfo {author} {\bibfnamefont {K.~G.}\ \bibnamefont
  {Wilson}}\ and\ \bibinfo {author} {\bibfnamefont {J.~B.}\ \bibnamefont
  {Kogut}},\ }\href {https://doi.org/10.1016/0370-1573(74)90023-4} {\bibfield
  {journal} {\bibinfo  {journal} {Phys.Rept.}\ }\textbf {\bibinfo {volume}
  {12}},\ \bibinfo {pages} {75} (\bibinfo {year} {1974})}\BibitemShut {NoStop}%
\bibitem [{\citenamefont
  {Symanzik}(1983{\natexlab{a}})}]{symanzik_continuum_1983b}%
  \BibitemOpen
  \bibfield  {author} {\bibinfo {author} {\bibfnamefont {K.}~\bibnamefont
  {Symanzik}},\ }\href {https://doi.org/10.1016/0550-3213(83)90468-6}
  {\bibfield  {journal} {\bibinfo  {journal} {Nuclear Physics B}\ }\textbf
  {\bibinfo {volume} {226}},\ \bibinfo {pages} {187} (\bibinfo {year}
  {1983}{\natexlab{a}})}\BibitemShut {NoStop}%
\bibitem [{\citenamefont
  {Symanzik}(1983{\natexlab{b}})}]{symanzik_continuum_1983a}%
  \BibitemOpen
  \bibfield  {author} {\bibinfo {author} {\bibfnamefont {K.}~\bibnamefont
  {Symanzik}},\ }\href {https://doi.org/10.1016/0550-3213(83)90469-8}
  {\bibfield  {journal} {\bibinfo  {journal} {Nuclear Physics B}\ }\textbf
  {\bibinfo {volume} {226}},\ \bibinfo {pages} {205} (\bibinfo {year}
  {1983}{\natexlab{b}})}\BibitemShut {NoStop}%
\bibitem [{\citenamefont {Kogut}\ and\ \citenamefont
  {Susskind}(1975)}]{kogut_hamiltonian_1975}%
  \BibitemOpen
  \bibfield  {author} {\bibinfo {author} {\bibfnamefont {J.}~\bibnamefont
  {Kogut}}\ and\ \bibinfo {author} {\bibfnamefont {L.}~\bibnamefont
  {Susskind}},\ }\href {https://doi.org/10.1103/PhysRevD.11.395} {\bibfield
  {journal} {\bibinfo  {journal} {Physical Review D}\ }\textbf {\bibinfo
  {volume} {11}},\ \bibinfo {pages} {395} (\bibinfo {year} {1975})}\BibitemShut
  {NoStop}%
\bibitem [{\citenamefont {Zohar}\ and\ \citenamefont
  {Burrello}(2015)}]{zohar_formulation_2015a}%
  \BibitemOpen
  \bibfield  {author} {\bibinfo {author} {\bibfnamefont {E.}~\bibnamefont
  {Zohar}}\ and\ \bibinfo {author} {\bibfnamefont {M.}~\bibnamefont
  {Burrello}},\ }\href {https://doi.org/10.1103/PhysRevD.91.054506} {\bibfield
  {journal} {\bibinfo  {journal} {Physical Review D}\ }\textbf {\bibinfo
  {volume} {91}},\ \bibinfo {pages} {054506} (\bibinfo {year}
  {2015})}\BibitemShut {NoStop}%
\bibitem [{\citenamefont {Zohar}\ \emph {et~al.}(2017)\citenamefont {Zohar},
  \citenamefont {Farace}, \citenamefont {Reznik},\ and\ \citenamefont
  {Cirac}}]{zohar_digital_2017}%
  \BibitemOpen
  \bibfield  {author} {\bibinfo {author} {\bibfnamefont {E.}~\bibnamefont
  {Zohar}}, \bibinfo {author} {\bibfnamefont {A.}~\bibnamefont {Farace}},
  \bibinfo {author} {\bibfnamefont {B.}~\bibnamefont {Reznik}},\ and\ \bibinfo
  {author} {\bibfnamefont {J.~I.}\ \bibnamefont {Cirac}},\ }\href
  {https://doi.org/10.1103/PhysRevA.95.023604} {\bibfield  {journal} {\bibinfo
  {journal} {Physical Review A}\ }\textbf {\bibinfo {volume} {95}},\ \bibinfo
  {pages} {023604} (\bibinfo {year} {2017})}\BibitemShut {NoStop}%
\bibitem [{\citenamefont {{NuQS Collaboration}}\ \emph
  {et~al.}(2019)\citenamefont {{NuQS Collaboration}}, \citenamefont
  {Alexandru}, \citenamefont {Bedaque}, \citenamefont {Lamm},\ and\
  \citenamefont {Lawrence}}]{nuqscollaboration_ensuremath_2019}%
  \BibitemOpen
  \bibfield  {author} {\bibinfo {author} {\bibnamefont {{NuQS Collaboration}}},
  \bibinfo {author} {\bibfnamefont {A.}~\bibnamefont {Alexandru}}, \bibinfo
  {author} {\bibfnamefont {P.~F.}\ \bibnamefont {Bedaque}}, \bibinfo {author}
  {\bibfnamefont {H.}~\bibnamefont {Lamm}},\ and\ \bibinfo {author}
  {\bibfnamefont {S.}~\bibnamefont {Lawrence}},\ }\href
  {https://doi.org/10.1103/PhysRevLett.123.090501} {\bibfield  {journal}
  {\bibinfo  {journal} {Physical Review Letters}\ }\textbf {\bibinfo {volume}
  {123}},\ \bibinfo {pages} {090501} (\bibinfo {year} {2019})}\BibitemShut
  {NoStop}%
\bibitem [{\citenamefont {Martinez}\ \emph {et~al.}(2016)\citenamefont
  {Martinez}, \citenamefont {Muschik}, \citenamefont {Schindler}, \citenamefont
  {Nigg}, \citenamefont {Erhard}, \citenamefont {Heyl}, \citenamefont {Hauke},
  \citenamefont {Dalmonte}, \citenamefont {Monz}, \citenamefont {Zoller},\ and\
  \citenamefont {Blatt}}]{martinez_realtime_2016a}%
  \BibitemOpen
  \bibfield  {author} {\bibinfo {author} {\bibfnamefont {E.~A.}\ \bibnamefont
  {Martinez}}, \bibinfo {author} {\bibfnamefont {C.~A.}\ \bibnamefont
  {Muschik}}, \bibinfo {author} {\bibfnamefont {P.}~\bibnamefont {Schindler}},
  \bibinfo {author} {\bibfnamefont {D.}~\bibnamefont {Nigg}}, \bibinfo {author}
  {\bibfnamefont {A.}~\bibnamefont {Erhard}}, \bibinfo {author} {\bibfnamefont
  {M.}~\bibnamefont {Heyl}}, \bibinfo {author} {\bibfnamefont {P.}~\bibnamefont
  {Hauke}}, \bibinfo {author} {\bibfnamefont {M.}~\bibnamefont {Dalmonte}},
  \bibinfo {author} {\bibfnamefont {T.}~\bibnamefont {Monz}}, \bibinfo {author}
  {\bibfnamefont {P.}~\bibnamefont {Zoller}},\ and\ \bibinfo {author}
  {\bibfnamefont {R.}~\bibnamefont {Blatt}},\ }\href
  {https://doi.org/10.1038/nature18318} {\bibfield  {journal} {\bibinfo
  {journal} {Nature}\ }\textbf {\bibinfo {volume} {534}},\ \bibinfo {pages}
  {516} (\bibinfo {year} {2016})}\BibitemShut {NoStop}%
\bibitem [{\citenamefont {Klco}\ and\ \citenamefont
  {Savage}(2019)}]{klco_digitization_2019}%
  \BibitemOpen
  \bibfield  {author} {\bibinfo {author} {\bibfnamefont {N.}~\bibnamefont
  {Klco}}\ and\ \bibinfo {author} {\bibfnamefont {M.~J.}\ \bibnamefont
  {Savage}},\ }\href {https://doi.org/10.1103/PhysRevA.99.052335} {\bibfield
  {journal} {\bibinfo  {journal} {Phys. Rev. A}\ }\textbf {\bibinfo {volume}
  {99}},\ \bibinfo {pages} {052335} (\bibinfo {year} {2019})}\BibitemShut
  {NoStop}%
\bibitem [{\citenamefont {Ji}\ \emph {et~al.}(2020)\citenamefont {Ji},
  \citenamefont {Lamm},\ and\ \citenamefont {Zhu}}]{ji_gluon_2020}%
  \BibitemOpen
  \bibfield  {author} {\bibinfo {author} {\bibfnamefont {Y.}~\bibnamefont
  {Ji}}, \bibinfo {author} {\bibfnamefont {H.}~\bibnamefont {Lamm}},\ and\
  \bibinfo {author} {\bibfnamefont {S.}~\bibnamefont {Zhu}},\ }\href
  {https://doi.org/10.1103/PhysRevD.102.114513} {\bibfield  {journal} {\bibinfo
   {journal} {Physical Review D}\ }\textbf {\bibinfo {volume} {102}},\ \bibinfo
  {pages} {114513} (\bibinfo {year} {2020})}\BibitemShut {NoStop}%
\bibitem [{\citenamefont {Haase}\ \emph {et~al.}(2021)\citenamefont {Haase},
  \citenamefont {Dellantonio}, \citenamefont {Celi}, \citenamefont {Paulson},
  \citenamefont {Kan}, \citenamefont {Jansen},\ and\ \citenamefont
  {Muschik}}]{haase_resource_2021}%
  \BibitemOpen
  \bibfield  {author} {\bibinfo {author} {\bibfnamefont {J.~F.}\ \bibnamefont
  {Haase}}, \bibinfo {author} {\bibfnamefont {L.}~\bibnamefont {Dellantonio}},
  \bibinfo {author} {\bibfnamefont {A.}~\bibnamefont {Celi}}, \bibinfo {author}
  {\bibfnamefont {D.}~\bibnamefont {Paulson}}, \bibinfo {author} {\bibfnamefont
  {A.}~\bibnamefont {Kan}}, \bibinfo {author} {\bibfnamefont {K.}~\bibnamefont
  {Jansen}},\ and\ \bibinfo {author} {\bibfnamefont {C.~A.}\ \bibnamefont
  {Muschik}},\ }\href {https://doi.org/10.22331/q-2021-02-04-393} {\bibfield
  {journal} {\bibinfo  {journal} {Quantum}\ }\textbf {\bibinfo {volume} {5}},\
  \bibinfo {pages} {393} (\bibinfo {year} {2021})}\BibitemShut {NoStop}%
\bibitem [{\citenamefont {Bruckmann}\ \emph {et~al.}(2019)\citenamefont
  {Bruckmann}, \citenamefont {Jansen},\ and\ \citenamefont
  {K{\"u}hn}}]{bruckmann_nonlinear_2019}%
  \BibitemOpen
  \bibfield  {author} {\bibinfo {author} {\bibfnamefont {F.}~\bibnamefont
  {Bruckmann}}, \bibinfo {author} {\bibfnamefont {K.}~\bibnamefont {Jansen}},\
  and\ \bibinfo {author} {\bibfnamefont {S.}~\bibnamefont {K{\"u}hn}},\ }\href
  {https://doi.org/10.1103/PhysRevD.99.074501} {\bibfield  {journal} {\bibinfo
  {journal} {Physical Review D}\ }\textbf {\bibinfo {volume} {99}},\ \bibinfo
  {pages} {074501} (\bibinfo {year} {2019})}\BibitemShut {NoStop}%
\bibitem [{\citenamefont {Bronzan}(1985)}]{bronzan_explicit_1985}%
  \BibitemOpen
  \bibfield  {author} {\bibinfo {author} {\bibfnamefont {J.~B.}\ \bibnamefont
  {Bronzan}},\ }\href {https://doi.org/10.1103/PhysRevD.31.2020} {\bibfield
  {journal} {\bibinfo  {journal} {Physical Review D}\ }\textbf {\bibinfo
  {volume} {31}},\ \bibinfo {pages} {2020} (\bibinfo {year}
  {1985})}\BibitemShut {NoStop}%
\bibitem [{\citenamefont {Bender}\ and\ \citenamefont
  {Zohar}(2020)}]{bender_gauge_2020}%
  \BibitemOpen
  \bibfield  {author} {\bibinfo {author} {\bibfnamefont {J.}~\bibnamefont
  {Bender}}\ and\ \bibinfo {author} {\bibfnamefont {E.}~\bibnamefont {Zohar}},\
  }\href {https://doi.org/10.1103/PhysRevD.102.114517} {\bibfield  {journal}
  {\bibinfo  {journal} {Physical Review D}\ }\textbf {\bibinfo {volume}
  {102}},\ \bibinfo {pages} {114517} (\bibinfo {year} {2020})}\BibitemShut
  {NoStop}%
\bibitem [{\citenamefont {Alexandru}\ \emph {et~al.}(2021)\citenamefont
  {Alexandru}, \citenamefont {Bedaque}, \citenamefont {Brett},\ and\
  \citenamefont {Lamm}}]{alexandru_spectrum_2021}%
  \BibitemOpen
  \bibfield  {author} {\bibinfo {author} {\bibfnamefont {A.}~\bibnamefont
  {Alexandru}}, \bibinfo {author} {\bibfnamefont {P.~F.}\ \bibnamefont
  {Bedaque}}, \bibinfo {author} {\bibfnamefont {R.}~\bibnamefont {Brett}},\
  and\ \bibinfo {author} {\bibfnamefont {H.}~\bibnamefont {Lamm}},\ }\href@noop
  {} {\bibfield  {journal} {\bibinfo  {journal} {arXiv:2112.08482 [hep-lat,
  physics:hep-ph, physics:quant-ph]}\ } (\bibinfo {year} {2021})}\BibitemShut
  {NoStop}%
\bibitem [{\citenamefont {Singh}\ and\ \citenamefont
  {Chandrasekharan}(2019{\natexlab{a}})}]{singh_qubit_2019}%
  \BibitemOpen
  \bibfield  {author} {\bibinfo {author} {\bibfnamefont {H.}~\bibnamefont
  {Singh}}\ and\ \bibinfo {author} {\bibfnamefont {S.}~\bibnamefont
  {Chandrasekharan}},\ }\href {https://doi.org/10.1103/PhysRevD.100.054505}
  {\bibfield  {journal} {\bibinfo  {journal} {Phys.Rev.}\ }\textbf {\bibinfo
  {volume} {D100}},\ \bibinfo {pages} {054505} (\bibinfo {year}
  {2019}{\natexlab{a}})}\BibitemShut {NoStop}%
\bibitem [{\citenamefont {Zhou}\ \emph {et~al.}(2021)\citenamefont {Zhou},
  \citenamefont {Singh}, \citenamefont {Bhattacharya}, \citenamefont
  {Chandrasekharan},\ and\ \citenamefont {Gupta}}]{zhou_spacetime_2021}%
  \BibitemOpen
  \bibfield  {author} {\bibinfo {author} {\bibfnamefont {J.}~\bibnamefont
  {Zhou}}, \bibinfo {author} {\bibfnamefont {H.}~\bibnamefont {Singh}},
  \bibinfo {author} {\bibfnamefont {T.}~\bibnamefont {Bhattacharya}}, \bibinfo
  {author} {\bibfnamefont {S.}~\bibnamefont {Chandrasekharan}},\ and\ \bibinfo
  {author} {\bibfnamefont {R.}~\bibnamefont {Gupta}},\ }\href@noop {}
  {\bibfield  {journal} {\bibinfo  {journal} {arxiv:2111.13780}\ } (\bibinfo
  {year} {2021})}\BibitemShut {NoStop}%
\bibitem [{\citenamefont {Bhattacharya}\ \emph {et~al.}(2021)\citenamefont
  {Bhattacharya}, \citenamefont {Buser}, \citenamefont {Chandrasekharan},
  \citenamefont {Gupta},\ and\ \citenamefont
  {Singh}}]{bhattacharya_qubit_2021}%
  \BibitemOpen
  \bibfield  {author} {\bibinfo {author} {\bibfnamefont {T.}~\bibnamefont
  {Bhattacharya}}, \bibinfo {author} {\bibfnamefont {A.~J.}\ \bibnamefont
  {Buser}}, \bibinfo {author} {\bibfnamefont {S.}~\bibnamefont
  {Chandrasekharan}}, \bibinfo {author} {\bibfnamefont {R.}~\bibnamefont
  {Gupta}},\ and\ \bibinfo {author} {\bibfnamefont {H.}~\bibnamefont {Singh}},\
  }\href {https://doi.org/10.1103/PhysRevLett.126.172001} {\bibfield  {journal}
  {\bibinfo  {journal} {Physical Review Letters}\ }\textbf {\bibinfo {volume}
  {126}},\ \bibinfo {pages} {172001} (\bibinfo {year} {2021})}\BibitemShut
  {NoStop}%
\bibitem [{\citenamefont {Liu}\ and\ \citenamefont
  {Chandrasekharan}(2022)}]{liu_qubit_2022}%
  \BibitemOpen
  \bibfield  {author} {\bibinfo {author} {\bibfnamefont {H.}~\bibnamefont
  {Liu}}\ and\ \bibinfo {author} {\bibfnamefont {S.}~\bibnamefont
  {Chandrasekharan}},\ }\href {https://doi.org/10.3390/sym14020305} {\bibfield
  {journal} {\bibinfo  {journal} {Symmetry}\ }\textbf {\bibinfo {volume}
  {14}},\ \bibinfo {pages} {305} (\bibinfo {year} {2022})}\BibitemShut
  {NoStop}%
\bibitem [{\citenamefont {Roy}\ \emph {et~al.}(2021)\citenamefont {Roy},
  \citenamefont {Schuricht}, \citenamefont {Hauschild}, \citenamefont
  {Pollmann},\ and\ \citenamefont {Saleur}}]{roy_quantum_2021}%
  \BibitemOpen
  \bibfield  {author} {\bibinfo {author} {\bibfnamefont {A.}~\bibnamefont
  {Roy}}, \bibinfo {author} {\bibfnamefont {D.}~\bibnamefont {Schuricht}},
  \bibinfo {author} {\bibfnamefont {J.}~\bibnamefont {Hauschild}}, \bibinfo
  {author} {\bibfnamefont {F.}~\bibnamefont {Pollmann}},\ and\ \bibinfo
  {author} {\bibfnamefont {H.}~\bibnamefont {Saleur}},\ }\href
  {https://doi.org/10.1016/j.nuclphysb.2021.115445} {\bibfield  {journal}
  {\bibinfo  {journal} {Nuclear Physics B}\ }\textbf {\bibinfo {volume}
  {968}},\ \bibinfo {pages} {115445} (\bibinfo {year} {2021})}\BibitemShut
  {NoStop}%
\bibitem [{\citenamefont {Roy}\ and\ \citenamefont
  {Saleur}(2019)}]{roy_quantum_2019}%
  \BibitemOpen
  \bibfield  {author} {\bibinfo {author} {\bibfnamefont {A.}~\bibnamefont
  {Roy}}\ and\ \bibinfo {author} {\bibfnamefont {H.}~\bibnamefont {Saleur}},\
  }\href {https://doi.org/10.1103/PhysRevB.100.155425} {\bibfield  {journal}
  {\bibinfo  {journal} {Physical Review B}\ }\textbf {\bibinfo {volume}
  {100}},\ \bibinfo {pages} {155425} (\bibinfo {year} {2019})}\BibitemShut
  {NoStop}%
\bibitem [{\citenamefont {{Zinn-Justin}}(2021)}]{zinn-justin_quantum_2021}%
  \BibitemOpen
  \bibfield  {author} {\bibinfo {author} {\bibfnamefont {J.}~\bibnamefont
  {{Zinn-Justin}}},\ }\href@noop {} {\emph {\bibinfo {title} {Quantum {{Field
  Theory}} and {{Critical Phenomena}}: {{Fifth Edition}}}}}\ (\bibinfo
  {publisher} {{Oxford University Press}},\ \bibinfo {year} {2021})\BibitemShut
  {NoStop}%
\bibitem [{\citenamefont {Campostrini}\ \emph {et~al.}(2006)\citenamefont
  {Campostrini}, \citenamefont {Hasenbusch}, \citenamefont {Pelissetto},\ and\
  \citenamefont {Vicari}}]{campostrini_critical_2006}%
  \BibitemOpen
  \bibfield  {author} {\bibinfo {author} {\bibfnamefont {M.}~\bibnamefont
  {Campostrini}}, \bibinfo {author} {\bibfnamefont {M.}~\bibnamefont
  {Hasenbusch}}, \bibinfo {author} {\bibfnamefont {A.}~\bibnamefont
  {Pelissetto}},\ and\ \bibinfo {author} {\bibfnamefont {E.}~\bibnamefont
  {Vicari}},\ }\href {https://doi.org/10.1103/PhysRevB.74.144506} {\bibfield
  {journal} {\bibinfo  {journal} {Phys.Rev.}\ }\textbf {\bibinfo {volume}
  {B74}},\ \bibinfo {pages} {144506} (\bibinfo {year} {2006})}\BibitemShut
  {NoStop}%
\bibitem [{\citenamefont {Pisarski}\ and\ \citenamefont
  {Wilczek}(1984)}]{pisarski_remarks_1984}%
  \BibitemOpen
  \bibfield  {author} {\bibinfo {author} {\bibfnamefont {R.~D.}\ \bibnamefont
  {Pisarski}}\ and\ \bibinfo {author} {\bibfnamefont {F.}~\bibnamefont
  {Wilczek}},\ }\href {https://doi.org/10.1103/PhysRevD.29.338} {\bibfield
  {journal} {\bibinfo  {journal} {Physical Review D}\ }\textbf {\bibinfo
  {volume} {29}},\ \bibinfo {pages} {338} (\bibinfo {year} {1984})}\BibitemShut
  {NoStop}%
\bibitem [{\citenamefont {Zamolodchikov}\ and\ \citenamefont
  {Zamolodchikov}(1979)}]{zamolodchikov_factorized_1979}%
  \BibitemOpen
  \bibfield  {author} {\bibinfo {author} {\bibfnamefont {A.~B.}\ \bibnamefont
  {Zamolodchikov}}\ and\ \bibinfo {author} {\bibfnamefont {A.~B.}\ \bibnamefont
  {Zamolodchikov}},\ }\href {https://doi.org/10.1016/0003-4916(79)90391-9}
  {\bibfield  {journal} {\bibinfo  {journal} {Annals of Physics}\ }\textbf
  {\bibinfo {volume} {120}},\ \bibinfo {pages} {253} (\bibinfo {year}
  {1979})}\BibitemShut {NoStop}%
\bibitem [{\citenamefont {Affleck}\ and\ \citenamefont
  {Haldane}(1987)}]{affleck_critical_1987a}%
  \BibitemOpen
  \bibfield  {author} {\bibinfo {author} {\bibfnamefont {I.}~\bibnamefont
  {Affleck}}\ and\ \bibinfo {author} {\bibfnamefont {F.~D.~M.}\ \bibnamefont
  {Haldane}},\ }\href {https://doi.org/10.1103/PhysRevB.36.5291} {\bibfield
  {journal} {\bibinfo  {journal} {Physical Review B}\ }\textbf {\bibinfo
  {volume} {36}},\ \bibinfo {pages} {5291} (\bibinfo {year}
  {1987})}\BibitemShut {NoStop}%
\bibitem [{\citenamefont
  {Haldane}(1983{\natexlab{a}})}]{haldane_continuum_1983}%
  \BibitemOpen
  \bibfield  {author} {\bibinfo {author} {\bibfnamefont {F.~D.~M.}\
  \bibnamefont {Haldane}},\ }\href
  {https://doi.org/10.1016/0375-9601(83)90631-X} {\bibfield  {journal}
  {\bibinfo  {journal} {Phys.Lett.}\ }\textbf {\bibinfo {volume} {A93}},\
  \bibinfo {pages} {464} (\bibinfo {year} {1983}{\natexlab{a}})}\BibitemShut
  {NoStop}%
\bibitem [{\citenamefont
  {Haldane}(1983{\natexlab{b}})}]{haldane_nonlinear_1983}%
  \BibitemOpen
  \bibfield  {author} {\bibinfo {author} {\bibfnamefont {F.~D.~M.}\
  \bibnamefont {Haldane}},\ }\href
  {https://doi.org/10.1103/PhysRevLett.50.1153} {\bibfield  {journal} {\bibinfo
   {journal} {Phys.Rev.Lett.}\ }\textbf {\bibinfo {volume} {50}},\ \bibinfo
  {pages} {1153} (\bibinfo {year} {1983}{\natexlab{b}})}\BibitemShut {NoStop}%
\bibitem [{\citenamefont {Wilson}\ and\ \citenamefont
  {Fisher}(1972)}]{wilson_critical_1972}%
  \BibitemOpen
  \bibfield  {author} {\bibinfo {author} {\bibfnamefont {K.~G.}\ \bibnamefont
  {Wilson}}\ and\ \bibinfo {author} {\bibfnamefont {M.~E.}\ \bibnamefont
  {Fisher}},\ }\href {https://doi.org/10.1103/PhysRevLett.28.240} {\bibfield
  {journal} {\bibinfo  {journal} {Physical Review Letters}\ }\textbf {\bibinfo
  {volume} {28}},\ \bibinfo {pages} {240} (\bibinfo {year} {1972})}\BibitemShut
  {NoStop}%
\bibitem [{\citenamefont {Cecile}\ and\ \citenamefont
  {Chandrasekharan}(2008)}]{cecile_modeling_2008}%
  \BibitemOpen
  \bibfield  {author} {\bibinfo {author} {\bibfnamefont {D.~J.}\ \bibnamefont
  {Cecile}}\ and\ \bibinfo {author} {\bibfnamefont {S.}~\bibnamefont
  {Chandrasekharan}},\ }\href {https://doi.org/10.1103/PhysRevD.77.014506}
  {\bibfield  {journal} {\bibinfo  {journal} {Phys.Rev.}\ }\textbf {\bibinfo
  {volume} {D77}},\ \bibinfo {pages} {014506} (\bibinfo {year}
  {2008})}\BibitemShut {NoStop}%
\bibitem [{\citenamefont {Banerjee}\ \emph {et~al.}(2018)\citenamefont
  {Banerjee}, \citenamefont {Chandrasekharan},\ and\ \citenamefont
  {Orlando}}]{banerjee_conformal_2018}%
  \BibitemOpen
  \bibfield  {author} {\bibinfo {author} {\bibfnamefont {D.}~\bibnamefont
  {Banerjee}}, \bibinfo {author} {\bibfnamefont {S.}~\bibnamefont
  {Chandrasekharan}},\ and\ \bibinfo {author} {\bibfnamefont {D.}~\bibnamefont
  {Orlando}},\ }\href {https://doi.org/10.1103/PhysRevLett.120.061603}
  {\bibfield  {journal} {\bibinfo  {journal} {Physical Review Letters}\
  }\textbf {\bibinfo {volume} {120}},\ \bibinfo {pages} {061603} (\bibinfo
  {year} {2018})}\BibitemShut {NoStop}%
\bibitem [{\citenamefont {Banerjee}\ \emph {et~al.}(2019)\citenamefont
  {Banerjee}, \citenamefont {Chandrasekharan}, \citenamefont {Orlando},\ and\
  \citenamefont {Reffert}}]{banerjee_conformal_2019}%
  \BibitemOpen
  \bibfield  {author} {\bibinfo {author} {\bibfnamefont {D.}~\bibnamefont
  {Banerjee}}, \bibinfo {author} {\bibfnamefont {S.}~\bibnamefont
  {Chandrasekharan}}, \bibinfo {author} {\bibfnamefont {D.}~\bibnamefont
  {Orlando}},\ and\ \bibinfo {author} {\bibfnamefont {S.}~\bibnamefont
  {Reffert}},\ }\href {https://doi.org/10.1103/PhysRevLett.123.051603}
  {\bibfield  {journal} {\bibinfo  {journal} {Phys.Rev.Lett.}\ }\textbf
  {\bibinfo {volume} {123}},\ \bibinfo {pages} {051603} (\bibinfo {year}
  {2019})}\BibitemShut {NoStop}%
\bibitem [{\citenamefont {Sylju{\aa}sen}\ and\ \citenamefont
  {Sandvik}(2002)}]{syljuasen_quantum_2002a}%
  \BibitemOpen
  \bibfield  {author} {\bibinfo {author} {\bibfnamefont {O.~F.}\ \bibnamefont
  {Sylju{\aa}sen}}\ and\ \bibinfo {author} {\bibfnamefont {A.~W.}\ \bibnamefont
  {Sandvik}},\ }\href {https://doi.org/10.1103/PhysRevE.66.046701} {\bibfield
  {journal} {\bibinfo  {journal} {Physical Review E}\ }\textbf {\bibinfo
  {volume} {66}},\ \bibinfo {pages} {046701} (\bibinfo {year}
  {2002})}\BibitemShut {NoStop}%
\bibitem [{\citenamefont {Singh}\ and\ \citenamefont
  {Chandrasekharan}(2019{\natexlab{b}})}]{singh_fewbody_2019}%
  \BibitemOpen
  \bibfield  {author} {\bibinfo {author} {\bibfnamefont {H.}~\bibnamefont
  {Singh}}\ and\ \bibinfo {author} {\bibfnamefont {S.}~\bibnamefont
  {Chandrasekharan}},\ }\href {https://doi.org/10.1103/PhysRevD.99.074511}
  {\bibfield  {journal} {\bibinfo  {journal} {Physical Review D}\ }\textbf
  {\bibinfo {volume} {99}},\ \bibinfo {pages} {074511} (\bibinfo {year}
  {2019}{\natexlab{b}})}\BibitemShut {NoStop}%
\bibitem [{\citenamefont {Beard}\ \emph {et~al.}(1998)\citenamefont {Beard},
  \citenamefont {Birgeneau}, \citenamefont {Greven},\ and\ \citenamefont
  {Wiese}}]{beard_squarelattice_1998a}%
  \BibitemOpen
  \bibfield  {author} {\bibinfo {author} {\bibfnamefont {B.~B.}\ \bibnamefont
  {Beard}}, \bibinfo {author} {\bibfnamefont {R.~J.}\ \bibnamefont
  {Birgeneau}}, \bibinfo {author} {\bibfnamefont {M.}~\bibnamefont {Greven}},\
  and\ \bibinfo {author} {\bibfnamefont {U.-J.}\ \bibnamefont {Wiese}},\ }\href
  {https://doi.org/10.1103/PhysRevLett.80.1742} {\bibfield  {journal} {\bibinfo
   {journal} {Physical Review Letters}\ }\textbf {\bibinfo {volume} {80}},\
  \bibinfo {pages} {1742} (\bibinfo {year} {1998})}\BibitemShut {NoStop}%
\bibitem [{\citenamefont {Beard}\ and\ \citenamefont
  {Wiese}(1996)}]{beard_simulations_1996}%
  \BibitemOpen
  \bibfield  {author} {\bibinfo {author} {\bibfnamefont {B.~B.}\ \bibnamefont
  {Beard}}\ and\ \bibinfo {author} {\bibfnamefont {U.-J.}\ \bibnamefont
  {Wiese}},\ }\href {https://doi.org/10.1103/PhysRevLett.77.5130} {\bibfield
  {journal} {\bibinfo  {journal} {Physical Review Letters}\ }\textbf {\bibinfo
  {volume} {77}},\ \bibinfo {pages} {5130} (\bibinfo {year}
  {1996})}\BibitemShut {NoStop}%
\bibitem [{\citenamefont {Beard}\ \emph {et~al.}(2005)\citenamefont {Beard},
  \citenamefont {Pepe}, \citenamefont {Riederer},\ and\ \citenamefont
  {Wiese}}]{beard_study_2005}%
  \BibitemOpen
  \bibfield  {author} {\bibinfo {author} {\bibfnamefont {B.~B.}\ \bibnamefont
  {Beard}}, \bibinfo {author} {\bibfnamefont {M.}~\bibnamefont {Pepe}},
  \bibinfo {author} {\bibfnamefont {S.}~\bibnamefont {Riederer}},\ and\
  \bibinfo {author} {\bibfnamefont {U.-J.}\ \bibnamefont {Wiese}},\ }\href
  {https://doi.org/10.1103/PhysRevLett.94.010603} {\bibfield  {journal}
  {\bibinfo  {journal} {Physical Review Letters}\ }\textbf {\bibinfo {volume}
  {94}},\ \bibinfo {pages} {010603} (\bibinfo {year} {2005})}\BibitemShut
  {NoStop}%
\bibitem [{\citenamefont {Prokof'ev}\ and\ \citenamefont
  {Svistunov}(2010)}]{prokofev_worm_2010}%
  \BibitemOpen
  \bibfield  {author} {\bibinfo {author} {\bibfnamefont {N.}~\bibnamefont
  {Prokof'ev}}\ and\ \bibinfo {author} {\bibfnamefont {B.}~\bibnamefont
  {Svistunov}},\ }\href@noop {} {\bibfield  {journal} {\bibinfo  {journal}
  {arXiv:0910.1393 [cond-mat, physics:hep-lat]}\ } (\bibinfo {year}
  {2010})}\BibitemShut {NoStop}%
\bibitem [{\citenamefont {Prokof'ev}\ and\ \citenamefont
  {Svistunov}(2001)}]{prokofev_worm_2001}%
  \BibitemOpen
  \bibfield  {author} {\bibinfo {author} {\bibfnamefont {N.}~\bibnamefont
  {Prokof'ev}}\ and\ \bibinfo {author} {\bibfnamefont {B.}~\bibnamefont
  {Svistunov}},\ }\href {https://doi.org/10.1103/PhysRevLett.87.160601}
  {\bibfield  {journal} {\bibinfo  {journal} {Phys.Rev.Lett.}\ }\textbf
  {\bibinfo {volume} {87}},\ \bibinfo {pages} {160601} (\bibinfo {year}
  {2001})}\BibitemShut {NoStop}%
\bibitem [{\citenamefont {Azcoiti}\ \emph {et~al.}(2009)\citenamefont
  {Azcoiti}, \citenamefont {Follana}, \citenamefont {Vaquero},\ and\
  \citenamefont {Carlo}}]{azcoiti_geometric_2009}%
  \BibitemOpen
  \bibfield  {author} {\bibinfo {author} {\bibfnamefont {V.}~\bibnamefont
  {Azcoiti}}, \bibinfo {author} {\bibfnamefont {E.}~\bibnamefont {Follana}},
  \bibinfo {author} {\bibfnamefont {A.}~\bibnamefont {Vaquero}},\ and\ \bibinfo
  {author} {\bibfnamefont {G.~D.}\ \bibnamefont {Carlo}},\ }\href
  {https://doi.org/10.1088/1126-6708/2009/08/008} {\bibfield  {journal}
  {\bibinfo  {journal} {Journal of High Energy Physics}\ }\textbf {\bibinfo
  {volume} {2009}},\ \bibinfo {pages} {008} (\bibinfo {year}
  {2009})}\BibitemShut {NoStop}%
\bibitem [{\citenamefont {Wolff}(2010)}]{wolff_simulating_2010}%
  \BibitemOpen
  \bibfield  {author} {\bibinfo {author} {\bibfnamefont {U.}~\bibnamefont
  {Wolff}},\ }\href {https://doi.org/10.1016/j.nuclphysb.2009.09.006}
  {\bibfield  {journal} {\bibinfo  {journal} {Nuclear Physics B}\ }\textbf
  {\bibinfo {volume} {824}},\ \bibinfo {pages} {254} (\bibinfo {year}
  {2010})}\BibitemShut {NoStop}%
\bibitem [{\citenamefont {Wolff}(2009{\natexlab{a}})}]{wolff_simulating_2009a}%
  \BibitemOpen
  \bibfield  {author} {\bibinfo {author} {\bibfnamefont {U.}~\bibnamefont
  {Wolff}},\ }\href {https://doi.org/10.1016/j.nuclphysb.2009.01.018}
  {\bibfield  {journal} {\bibinfo  {journal} {Nuclear Physics B}\ }\textbf
  {\bibinfo {volume} {814}},\ \bibinfo {pages} {549} (\bibinfo {year}
  {2009}{\natexlab{a}})}\BibitemShut {NoStop}%
\bibitem [{\citenamefont {Wolff}(2009{\natexlab{b}})}]{wolff_simulating_2009}%
  \BibitemOpen
  \bibfield  {author} {\bibinfo {author} {\bibfnamefont {U.}~\bibnamefont
  {Wolff}},\ }\href {https://doi.org/10.1016/j.nuclphysb.2008.09.033}
  {\bibfield  {journal} {\bibinfo  {journal} {Nuclear Physics B}\ }\textbf
  {\bibinfo {volume} {810}},\ \bibinfo {pages} {491} (\bibinfo {year}
  {2009}{\natexlab{b}})}\BibitemShut {NoStop}%
\bibitem [{\citenamefont {Wenger}(2009)}]{wenger_efficient_2009}%
  \BibitemOpen
  \bibfield  {author} {\bibinfo {author} {\bibfnamefont {U.}~\bibnamefont
  {Wenger}},\ }\href {https://doi.org/10.1103/PhysRevD.80.071503} {\bibfield
  {journal} {\bibinfo  {journal} {Physical Review D}\ }\textbf {\bibinfo
  {volume} {80}},\ \bibinfo {pages} {071503} (\bibinfo {year}
  {2009})}\BibitemShut {NoStop}%
\bibitem [{\citenamefont {Huffman}\ and\ \citenamefont
  {Chandrasekharan}(2017)}]{huffman_fermion_2017}%
  \BibitemOpen
  \bibfield  {author} {\bibinfo {author} {\bibfnamefont {E.}~\bibnamefont
  {Huffman}}\ and\ \bibinfo {author} {\bibfnamefont {S.}~\bibnamefont
  {Chandrasekharan}},\ }\href {https://doi.org/10.1103/PhysRevD.96.114502}
  {\bibfield  {journal} {\bibinfo  {journal} {Phys.Rev.}\ }\textbf {\bibinfo
  {volume} {D96}},\ \bibinfo {pages} {114502} (\bibinfo {year}
  {2017})}\BibitemShut {NoStop}%
\bibitem [{\citenamefont
  {Chandrasekharan}(2010)}]{chandrasekharan_fermion_2010}%
  \BibitemOpen
  \bibfield  {author} {\bibinfo {author} {\bibfnamefont {S.}~\bibnamefont
  {Chandrasekharan}},\ }\href {https://doi.org/10.1103/PhysRevD.82.025007}
  {\bibfield  {journal} {\bibinfo  {journal} {Physical Review D}\ }\textbf
  {\bibinfo {volume} {82}},\ \bibinfo {pages} {025007} (\bibinfo {year}
  {2010})}\BibitemShut {NoStop}%
\bibitem [{\citenamefont {Antonenko}\ and\ \citenamefont
  {Sokolov}(1995)}]{antonenko_critical_1995}%
  \BibitemOpen
  \bibfield  {author} {\bibinfo {author} {\bibfnamefont {S.~A.}\ \bibnamefont
  {Antonenko}}\ and\ \bibinfo {author} {\bibfnamefont {A.~I.}\ \bibnamefont
  {Sokolov}},\ }\href {https://doi.org/10.1103/PhysRevE.51.1894} {\bibfield
  {journal} {\bibinfo  {journal} {Physical Review E}\ }\textbf {\bibinfo
  {volume} {51}},\ \bibinfo {pages} {1894} (\bibinfo {year}
  {1995})}\BibitemShut {NoStop}%
\bibitem [{\citenamefont {Moshe}\ and\ \citenamefont
  {{Zinn-Justin}}(2003)}]{moshe_quantum_2003}%
  \BibitemOpen
  \bibfield  {author} {\bibinfo {author} {\bibfnamefont {M.}~\bibnamefont
  {Moshe}}\ and\ \bibinfo {author} {\bibfnamefont {J.}~\bibnamefont
  {{Zinn-Justin}}},\ }\href {https://doi.org/10.1016/S0370-1573(03)00263-1}
  {\bibfield  {journal} {\bibinfo  {journal} {Phys.Rept.}\ }\textbf {\bibinfo
  {volume} {385}},\ \bibinfo {pages} {69} (\bibinfo {year} {2003})}\BibitemShut
  {NoStop}%
\bibitem [{\citenamefont {Chester}\ \emph {et~al.}(2020)\citenamefont
  {Chester}, \citenamefont {Landry}, \citenamefont {Liu}, \citenamefont
  {Poland}, \citenamefont {{Simmons-Duffin}}, \citenamefont {Su},\ and\
  \citenamefont {Vichi}}]{chester_carving_2020}%
  \BibitemOpen
  \bibfield  {author} {\bibinfo {author} {\bibfnamefont {S.~M.}\ \bibnamefont
  {Chester}}, \bibinfo {author} {\bibfnamefont {W.}~\bibnamefont {Landry}},
  \bibinfo {author} {\bibfnamefont {J.}~\bibnamefont {Liu}}, \bibinfo {author}
  {\bibfnamefont {D.}~\bibnamefont {Poland}}, \bibinfo {author} {\bibfnamefont
  {D.}~\bibnamefont {{Simmons-Duffin}}}, \bibinfo {author} {\bibfnamefont
  {N.}~\bibnamefont {Su}},\ and\ \bibinfo {author} {\bibfnamefont
  {A.}~\bibnamefont {Vichi}},\ }\href {https://doi.org/10.1007/JHEP06(2020)142}
  {\bibfield  {journal} {\bibinfo  {journal} {Journal of High Energy Physics}\
  }\textbf {\bibinfo {volume} {2020}},\ \bibinfo {pages} {142} (\bibinfo {year}
  {2020})}\BibitemShut {NoStop}%
\bibitem [{\citenamefont {Deng}(2006)}]{deng_bulk_2006}%
  \BibitemOpen
  \bibfield  {author} {\bibinfo {author} {\bibfnamefont {Y.}~\bibnamefont
  {Deng}},\ }\href {https://doi.org/10.1103/PhysRevE.73.056116} {\bibfield
  {journal} {\bibinfo  {journal} {Physical Review E}\ }\textbf {\bibinfo
  {volume} {73}},\ \bibinfo {pages} {056116} (\bibinfo {year}
  {2006})}\BibitemShut {NoStop}%
\bibitem [{\citenamefont {Butti}\ and\ \citenamefont
  {Parisen~Toldin}(2005)}]{butti_critical_2005}%
  \BibitemOpen
  \bibfield  {author} {\bibinfo {author} {\bibfnamefont {A.}~\bibnamefont
  {Butti}}\ and\ \bibinfo {author} {\bibfnamefont {F.}~\bibnamefont
  {Parisen~Toldin}},\ }\href {https://doi.org/10.1016/j.nuclphysb.2004.10.021}
  {\bibfield  {journal} {\bibinfo  {journal} {Nuclear Physics B}\ }\textbf
  {\bibinfo {volume} {704}},\ \bibinfo {pages} {527} (\bibinfo {year}
  {2005})}\BibitemShut {NoStop}%
\bibitem [{\citenamefont {Holtmann}\ and\ \citenamefont
  {Schulze}(2003)}]{holtmann_critical_2003}%
  \BibitemOpen
  \bibfield  {author} {\bibinfo {author} {\bibfnamefont {S.}~\bibnamefont
  {Holtmann}}\ and\ \bibinfo {author} {\bibfnamefont {T.}~\bibnamefont
  {Schulze}},\ }\href {https://doi.org/10.1103/PhysRevE.68.036111} {\bibfield
  {journal} {\bibinfo  {journal} {Physical Review E}\ }\textbf {\bibinfo
  {volume} {68}},\ \bibinfo {pages} {036111} (\bibinfo {year}
  {2003})}\BibitemShut {NoStop}%
\bibitem [{\citenamefont {Hasenbusch}\ and\ \citenamefont
  {Vicari}(2011)}]{hasenbusch_anisotropic_2011}%
  \BibitemOpen
  \bibfield  {author} {\bibinfo {author} {\bibfnamefont {M.}~\bibnamefont
  {Hasenbusch}}\ and\ \bibinfo {author} {\bibfnamefont {E.}~\bibnamefont
  {Vicari}},\ }\href {https://doi.org/10.1103/PhysRevB.84.125136} {\bibfield
  {journal} {\bibinfo  {journal} {Physical Review B}\ }\textbf {\bibinfo
  {volume} {84}},\ \bibinfo {pages} {125136} (\bibinfo {year}
  {2011})}\BibitemShut {NoStop}%
\bibitem [{\citenamefont {Hasenbusch}(2019)}]{hasenbusch_monte_2019}%
  \BibitemOpen
  \bibfield  {author} {\bibinfo {author} {\bibfnamefont {M.}~\bibnamefont
  {Hasenbusch}},\ }\href {https://doi.org/10.1103/PhysRevB.100.224517}
  {\bibfield  {journal} {\bibinfo  {journal} {Physical Review B}\ }\textbf
  {\bibinfo {volume} {100}},\ \bibinfo {pages} {224517} (\bibinfo {year}
  {2019})}\BibitemShut {NoStop}%
\bibitem [{\citenamefont {Pelissetto}\ and\ \citenamefont
  {Vicari}(2002)}]{pelissetto_critical_2002}%
  \BibitemOpen
  \bibfield  {author} {\bibinfo {author} {\bibfnamefont {A.}~\bibnamefont
  {Pelissetto}}\ and\ \bibinfo {author} {\bibfnamefont {E.}~\bibnamefont
  {Vicari}},\ }\href {https://doi.org/10.1016/S0370-1573(02)00219-3} {\bibfield
   {journal} {\bibinfo  {journal} {Phys.Rept.}\ }\textbf {\bibinfo {volume}
  {368}},\ \bibinfo {pages} {549} (\bibinfo {year} {2002})}\BibitemShut
  {NoStop}%
\bibitem [{\citenamefont {Kos}\ \emph {et~al.}(2016)\citenamefont {Kos},
  \citenamefont {Poland}, \citenamefont {{Simmons-Duffin}},\ and\ \citenamefont
  {Vichi}}]{kos_precision_2016}%
  \BibitemOpen
  \bibfield  {author} {\bibinfo {author} {\bibfnamefont {F.}~\bibnamefont
  {Kos}}, \bibinfo {author} {\bibfnamefont {D.}~\bibnamefont {Poland}},
  \bibinfo {author} {\bibfnamefont {D.}~\bibnamefont {{Simmons-Duffin}}},\ and\
  \bibinfo {author} {\bibfnamefont {A.}~\bibnamefont {Vichi}},\ }\bibfield
  {journal} {\bibinfo  {journal} {Journal of High Energy Physics}\ }\textbf
  {\bibinfo {volume} {2016}},\ \href {https://doi.org/10.1007/JHEP08(2016)036}
  {10.1007/JHEP08(2016)036} (\bibinfo {year} {2016})\BibitemShut {NoStop}%
\bibitem [{\citenamefont {Chandrasekharan}\ and\ \citenamefont
  {Wiese}(1997)}]{chandrasekharan_quantum_1997}%
  \BibitemOpen
  \bibfield  {author} {\bibinfo {author} {\bibfnamefont {S.}~\bibnamefont
  {Chandrasekharan}}\ and\ \bibinfo {author} {\bibfnamefont {U.~J.}\
  \bibnamefont {Wiese}},\ }\href
  {https://doi.org/10.1016/S0550-3213(97)00006-0} {\bibfield  {journal}
  {\bibinfo  {journal} {Nucl.Phys.}\ }\textbf {\bibinfo {volume} {B492}},\
  \bibinfo {pages} {455} (\bibinfo {year} {1997})}\BibitemShut {NoStop}%
\bibitem [{\citenamefont {Brower}\ \emph {et~al.}(1999)\citenamefont {Brower},
  \citenamefont {Chandrasekharan},\ and\ \citenamefont
  {Wiese}}]{brower_qcd_1999}%
  \BibitemOpen
  \bibfield  {author} {\bibinfo {author} {\bibfnamefont {R.}~\bibnamefont
  {Brower}}, \bibinfo {author} {\bibfnamefont {S.}~\bibnamefont
  {Chandrasekharan}},\ and\ \bibinfo {author} {\bibfnamefont {U.~J.}\
  \bibnamefont {Wiese}},\ }\href {https://doi.org/10.1103/PhysRevD.60.094502}
  {\bibfield  {journal} {\bibinfo  {journal} {Phys.Rev.}\ }\textbf {\bibinfo
  {volume} {D60}},\ \bibinfo {pages} {094502} (\bibinfo {year}
  {1999})}\BibitemShut {NoStop}%
\bibitem [{\citenamefont {Chandrasekharan}\ \emph {et~al.}(2002)\citenamefont
  {Chandrasekharan}, \citenamefont {Scarlet},\ and\ \citenamefont
  {Wiese}}]{chandrasekharan_spin_2002}%
  \BibitemOpen
  \bibfield  {author} {\bibinfo {author} {\bibfnamefont {S.}~\bibnamefont
  {Chandrasekharan}}, \bibinfo {author} {\bibfnamefont {B.}~\bibnamefont
  {Scarlet}},\ and\ \bibinfo {author} {\bibfnamefont {U.~J.}\ \bibnamefont
  {Wiese}},\ }\href {https://doi.org/10.1016/S0010-4655(02)00311-9} {\bibfield
  {journal} {\bibinfo  {journal} {Computer Physics Communications}\ }\bibinfo
  {series} {Proceedings of the {{Europhysics Conference}} on {{Computational
  Physics Computational Modeling}} and {{Simulation}} of {{Complex Systems}}},\
  \textbf {\bibinfo {volume} {147}},\ \bibinfo {pages} {388} (\bibinfo {year}
  {2002})}\BibitemShut {NoStop}%
\bibitem [{\citenamefont {Brower}\ \emph {et~al.}(2004)\citenamefont {Brower},
  \citenamefont {Chandrasekharan}, \citenamefont {Riederer},\ and\
  \citenamefont {Wiese}}]{brower_dtheory_2004}%
  \BibitemOpen
  \bibfield  {author} {\bibinfo {author} {\bibfnamefont {R.}~\bibnamefont
  {Brower}}, \bibinfo {author} {\bibfnamefont {S.}~\bibnamefont
  {Chandrasekharan}}, \bibinfo {author} {\bibfnamefont {S.}~\bibnamefont
  {Riederer}},\ and\ \bibinfo {author} {\bibfnamefont {U.~J.}\ \bibnamefont
  {Wiese}},\ }\href {https://doi.org/10.1016/j.nuclphysb.2004.06.007}
  {\bibfield  {journal} {\bibinfo  {journal} {Nuclear Physics B}\ }\textbf
  {\bibinfo {volume} {693}},\ \bibinfo {pages} {149} (\bibinfo {year}
  {2004})}\BibitemShut {NoStop}%
\bibitem [{\citenamefont {Niedermayer}\ and\ \citenamefont
  {Wolff}(2016)}]{niedermayer_model_2016}%
  \BibitemOpen
  \bibfield  {author} {\bibinfo {author} {\bibfnamefont {F.}~\bibnamefont
  {Niedermayer}}\ and\ \bibinfo {author} {\bibfnamefont {U.}~\bibnamefont
  {Wolff}},\ }\href {https://doi.org/10.22323/1.256.0317} {\bibfield  {journal}
  {\bibinfo  {journal} {PoS}\ }\textbf {\bibinfo {volume} {LATTICE2016}},\
  \bibinfo {pages} {317} (\bibinfo {year} {2016})}\BibitemShut {NoStop}%
\end{thebibliography}%

\appendix
\section{Fits to the scaling behavior} 
\label{sec:fits-scaling}

Close to a critical point, physical observables exhibit scaling behavior.
In particular,  the two observables current-current susceptibility $\rho_s$ and susceptibility $\chi$,  defined above in \cref{sec:qubit-observables} should  behave as
\begin{align}
    \rho_s(J; L) &= \frac{1}{L^{D-2}} f(u) \label{eq:scaling-behavior-rho}\\
    \chi(J; L) &= {L^{2-\eta}} g(u),
              \label{eq:scaling-behavior-chi}
\end{align}
where $D$ is the spacetime dimension,
$u=(J-J_c) L^{1/\nu}$ is the scaling variable,
$f(u)$ and $g(u)$ are unknown universal scaling functions analytic at $u=0$,
and $\eta$ and $\nu$ are the critical exponents
From the data for these observables $\rho_s$ and $\chi$ at various box sizes $L$ and couplings $J$ close to the critical value $J_c$, we can perform a combined fit to \cref{eq:scaling-behavior-rho,eq:scaling-behavior-chi}.
We approximate the functions $f(u)$ and $g(u)$ by a truncated Taylor
series expansion up to a given order $\nMax$
\begin{align}
  f(u) &= \sum_{n=0}^{\nMax}f_n u^{n}, \\
  g(u) &= \sum_{n=0}^{\nMax}g_n u^{n}.
\end{align}
The coefficients $f_i$, $g_i$ become  parameters to be fit.
Since $u$ is small close to the critical point, the Taylor series can be truncated to small values such as $\nMax=2,3,4$.
However, we do need to vary $\nMax$ empirically to find an optimal truncation.
This procedure allows us to get a fairly accurate first estimate for the critical coupling $J_c$, the critical exponents $\nu$, $\eta$, and the scaling functions $f(u)$ and $g(u)$.

However, this nonlinear fitting procedure (with $2\nMax + 5$ parameters) is somewhat opaque given its reliance on the correct range of $L$ and $J$, good initial guesses for the parameters and optimal polynomial truncations for the functions $f(u)$ and $g(u)$. Hence, we perform a few additional steps to make sure that we are indeed in the scaling regime, and to identify the critical exponents more precisely.

\begin{figure*}[p]
    \centering
    \def\PlotWidthFactor{0.45}
    \includegraphics[width=\PlotWidthFactor\textwidth]{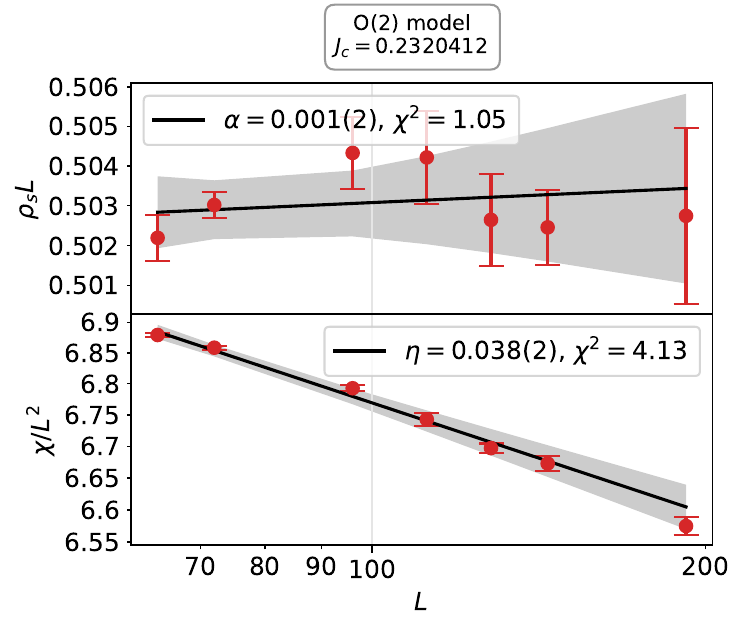}
    \includegraphics[width=\PlotWidthFactor\textwidth]{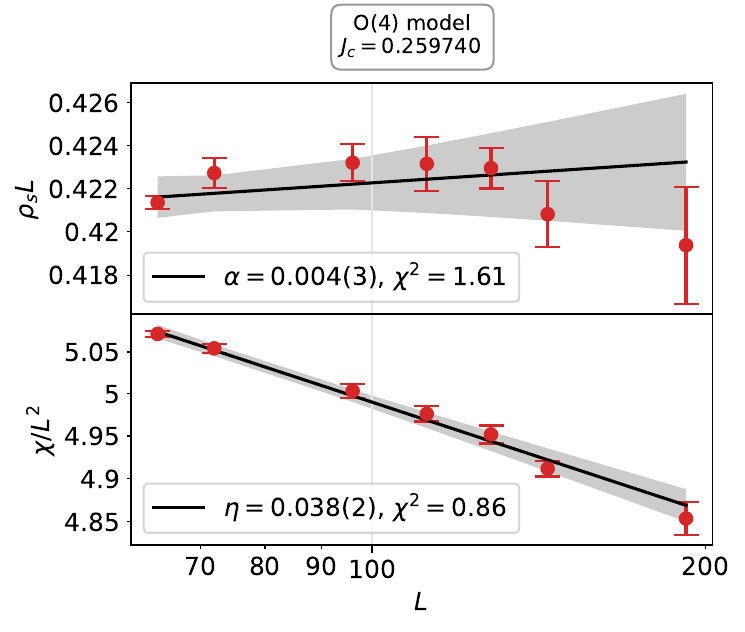}\\
    \includegraphics[width=\PlotWidthFactor\textwidth]{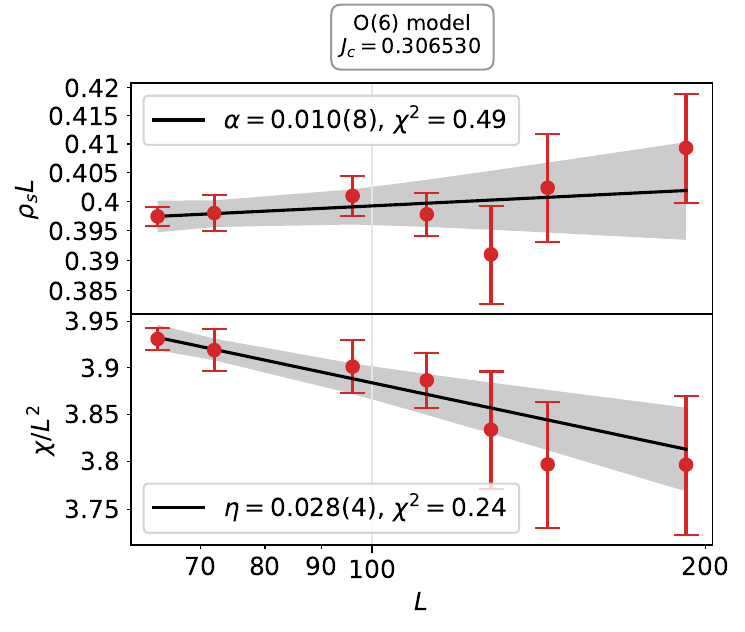}
    \includegraphics[width=\PlotWidthFactor\textwidth]{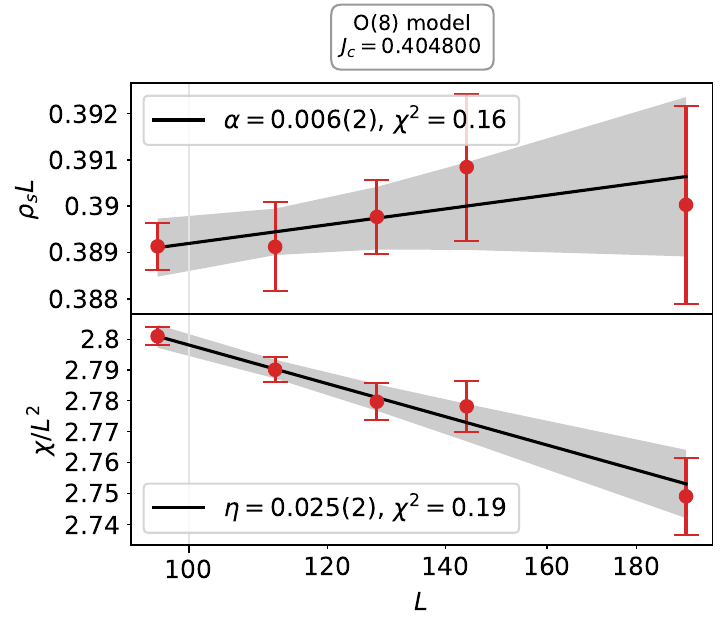}
    \caption{
      Critical scaling for $O(N)$  models with $N=2,\dotsc,8$ at the critical point $J=J_c$.
      The top plot for each $N$: We perform a powerlaw fit for winding-number susceptibility $\rho_s L = f(0) L^\alpha$ in a range of box sizes $[\Lmin, \Lmax]$ and extract the power $\alpha$.
      If the box sizes are large enough, then we must have $\alpha=0$ at the critical points.
      By choosing the window $[\Lmin, \Lmax]$ such that $\alpha \approx 0$, we can make sure our subsequent fits are in the scaling regime.
      The bottom plot for each $N$: We then fit the susceptibility to the form $\chi(0; L) = g(0) L^{2-\eta}$ to extract $\eta$.
      We can now use this value of $\eta$, as well as $f(0), g(0)$ as an input during the extraction of $\nu$ from the double fits, as demonstrated in \cref{fig:critical-scaling-1}.
      }
    \label{fig:critical-scaling-2}
\end{figure*}

Once we obtain a precise estimate for the critical point $J=J_c$ from the above fits, we perform another set of computations exactly at $J=J_c$ for various $L$, shown in \cref{fig:critical-scaling-2}. First, we look at the current-current susceptibility $\rho_s(0; L)$ at $J=J_c$, which according to \cref{eq:scaling-behavior-rho}, must simply behave as
\begin{align}
    \rho_s(0; L) = \frac{f(0)}{L}.
  \label{eq:rho-critical}
\end{align}
for $D=3$ spacetime dimensions.
We perform a single powerlaw fit to  $\rho_s(0;L)L = f(0) L^{\alpha}$ to find $f(0)$ and the power $\alpha$.
If we are exactly at the critical point, the exponent must be $\alpha=0$, within errors.
So, we repeat this computation over a range of box sizes $L \in [\Lmin, \Lmax]$, until $\alpha$ becomes consistent with $\alpha=0$ within errors.
This gives us the scaling window $[\Lmin, \Lmax]$. These fits are shown in the top plots for each $N$ in \cref{fig:critical-scaling-2}.

Performing the computations at $J=J_c$ also allows for a clean extraction of the critical exponent $\eta$.
We perform a power-law fit of the susceptibility $\chi(0; L)$ to the form
\begin{align}
    \chi(0; L) = {g(0)}{L^{2-\eta}},
  \label{eq:chi-critical}
\end{align}
obtained by setting $u=0$ in \cref{eq:scaling-behavior-chi}.
This gives us our final value of the critical exponent $\eta$.
The bottom plot for each $N$ in  \cref{fig:critical-scaling-2}
shows the extraction of $\eta$ at the critical point.

Now we need to find the critical exponent $\nu$. 
For this, we go back and perform the combined fits of \cref{eq:scaling-behavior-rho,eq:scaling-behavior-chi} again.
But this time we use the scaling window $L=[\Lmin, \Lmax]$ determined above.
We also keep $\eta$, $f(0)$, and $g(0)$ fixed to the values obtained from the powerlaw fits at $J=J_c$ in \cref{eq:chi-critical,eq:rho-critical}.
This allows us to obtain a more controlled fit, and a more precise extraction of the critical exponent $\nu$.
This is the value for $\nu$ that we finally report.
Additionally, we obtain the critical point $J_c$, again to make sure it is consistent with the value we chose in the previous step.
This is an additional check of self-consistency of our fits.
\Cref{fig:critical-scaling-1} shows these combined fits in the scaling window and the extraction of the critical exponent $\nu$, the critical point $J_c$, and the universal scaling functions $f(u)$ and $g(u)$.

\end{document}